\documentclass{article}
\usepackage{emulateapj}

\lefthead{Nelson et al}
\righthead{Revisiting BCG Evolution with the LCDCS}
\newcommand\Kp{$K^\prime$\ } \newcommand\etal{\textit{et al.\ }}
 
\newcommand\msun{{\rm\,M_\odot}}

\def\msun{{\rm\,M_\odot}} 
 
 \def\Kp{$K^\prime$\ } \def\etal{{\it et
al.\ }}  

\begin{document}

\title{Revisiting Brightest Cluster Galaxy Evolution with the Las
Campanas Distant Cluster Survey} 
\author{Amy E. Nelson} \affil{Board of Astronomy and Astrophysics,
Univ. of California, Santa Cruz, CA, 95064, E-Mail: anelson@ucolick.org}
\author{Anthony H. Gonzalez}
\affil{Dept. of Astronomy, University of Florida, Gainesville, Fl, 32611, E-Mail: anthony@astro.ufl.edu} 
\author{Dennis Zaritsky} \affil{Steward
Observatory, 933 N. Cherry Ave., Univ. of Arizona, Tucson, AZ, 85721,
E-Mail: dzaritsky@as.arizona.edu} 
\author{and} 
\author{Julianne J. Dalcanton} \affil{Box 351580, University of
Washington, Seattle, WA, 98195, E-Mail:
jd@astro.washington.edu}

\begin{abstract}

We investigate the influence of environment on brightest cluster
galaxy (BCG) evolution using a sample of 63 clusters at 0.3 $\leq z
\leq$ 0.9 drawn primarily from the Las Campanas Distant Cluster Survey
and follow-up $V$, $I$, and \Kp photometry. The luminosity evolution
of the entire BCG sample is not adequately described by a single
evolutionary model.  Using the integrated light from the cluster
detection as a proxy for cluster L$_{x}$ and
the suggestion by Burke, Collins, \& Mann, we set L$_{x} = 2 \times
10^{44}$ ergs s$^{-1}$ to be the division between high and low
luminosity clusters.  At high redshift ($z>0.6$) BCGs from low-L$_{x}$
clusters are fainter, on average, than those from high-L$_{x}$
clusters and are best modeled as having constant luminosity with
redshift. The BCGs from high-L$_{x}$ cluster are best modeled as
having a stellar population that formed at large redshift
($z_{form}>5$) and is passively evolving. However, for the entire BCG
population, the observed $V-I$ and $I-$\Kp colors are well described
by a single evolutionary model in which the stellar populations have
$z_{form} > 5$ and subsequently passively evolve.  We conclude that
accretion is proportionally more significant for BCGs in lower mass
clusters at these redshifts (factor of 2 to 4 increase in mass since
$z \sim 1$ for the low L$_x$ systems; Aragon-Salamanca \etal) and that
the accreted matter is in the form of systems with evolved stellar
populations.

\end{abstract}


\section{Introduction}

In standard hierarchical formation scenarios, structure formation
occurs first in regions of highest overdensity. As a result of this
accelerated formation, it is generally believed, for example, that
cluster galaxies are on average older than field galaxies.  However,
studies comparing the evolution of galaxies in different environments
are difficult because identifying equivalent field and cluster
populations is an ill-defined problem and because the effects of
various physical processes (interactions, ram pressure stripping,
tidally triggered star formation, harassment by the global cluster
potential, etc.) are poorly understood. Selection by morphology,
luminosity, color, environment, or any other observable property of
galaxies has the potential to bias the sample.

One galaxy population that can be unambiguously identified in a
variety of environments, by definition, is the dominant
galaxy, referred to as the brightest cluster galaxy (BCG) in clusters
and the brightest group galaxy (BGG) in groups. Although the
particular galaxy that is the BCG or BGG may change with time in any
environment, that process is part of the evolution one aims to study
and the sample definition is relatively unambiguous and well-suited
for comparison to simulations. Over the last 50 years, many have
studied the properties of BCGs from redshifts 0 to $\sim$ 1 (some
highlights include Humason, Mayall, \& Sandage 1956; Sandage,
Kristian, \& Westphall 1976; Hoessel 1980; Schombert 1986; Graham \etal 1996); initially with the intent to use BCGs as
standard candles, later to study galaxy evolution.  We revisit the
topic of BCG evolution in light of recent claims of significant
accretion (Aragon-Salamanca, Baugh, \& Kauffmann 1998; hereafter ABK)
and environmentally induced differences (Burke, Collins, \& Mann 2000;
hereafter BCM) using a large sample of BCGs drawn from the Las
Campanas Distant Cluster Survey (LCDCS; Gonzalez \etal
2001a). Although BCG evolution may be quite distinct from galaxy
evolution in general, its study may shed light on the process of
hierarchical structure formation in dense environments.

BCGs are the most massive and luminous galaxies. However, the findings
that at least some BCGs are not drawn from the luminosity function of
cluster ellipticals (Tremaine \& Richstone 1977; Dressler 1978) and
that BCGs may have different luminosity profiles than other giant
ellipticals (Oemler 1976; Schombert 1986) suggest that they have a
different evolutionary history than standard cluster
ellipticals. Aragon-Salamanca \etal (1993; hereafter A93) observed
BCGs at high redshift ($0.5 < z < 0.9$) in the $K$-band and found that
while the BCG colors agree with the prediction of passive evolution
models with high formation redshifts ($z_{form} \gtrsim 2$), similar
to the results for cluster ellipticals (Bower, Lucey, \& Ellis 1992),
their luminosities match the prediction of no-evolution
models. Because stellar populations dim as they age, ABK interpret
this lack of even passive luminosity evolution to imply significant,
relatively recent mass accretion by the BCGs.  Parameterizing the
necessary mass accretion, ABK estimate that BCGs have grown by a
factor of two to four in mass since $z \sim 1$, a result which they
find to be consistent with the expectation from hierarchical models as
described by semi-analytic models of galaxy evolution (cf. Kauffmann,
White \& Guiderdoni 1993)

This rather attractive agreement between observations and simulations
for the evolution of BCGs over $\sim$ 75\% of the age of the Universe
has recently been questioned. The focus of the current discussion is
whether all BCGs are similar, and in particular whether environment
plays a critical role in determining the properties of BCGs (Hudson \&
Ebeling 1997; Collins \& Mann 1998, hereafter CM; and BCM). BCM study
the BCGs of 76 x-ray selected clusters in the $K$-band and find that
BCGs from low-L$_{x}$ clusters (L$_{x} < 2.3 \times 10^{44}$ ergs
s$^{-1}$) display no luminosity evolution.  (We will refer to the BCGs
from high and low L$_x$ clusters as high-L$_{x}$ and low-L$_{x}$
BCGs.) Furthermore, they find that the ABK sample primarily consists
of low-L$_{x}$ BCGs at $z>0.5$. Therefore, the BCM observation of no
luminosity evolution in the BCGs from low-L$_{x}$ clusters is entirely
consistent with that of ABK. However, they find that the luminosities
of high-L$_{x}$ BCGs have evolved since $z \sim 1$. The stellar
populations of their high-L$_{x}$
BCGs are consistent with the luminosity predictions of passive
evolution models with a high formation redshift ($z_{form} \gtrsim
5$). Because high-L$_{x}$ clusters best match the cluster selection
adopted for comparison to the semi-analytic models, they reinvestigate
the parameterization of the mass evolution of high-L$_{x}$ BCGs and
find substantially lower rates of mass accretion, at most a factor of
$\sim$2 increase in mass since $z \sim 1$.  This result suggests that
BCGs from high-L$_{x}$ clusters experience a different evolutionary
scenario than those from low-L$_{x}$ clusters.

Because BCG properties appear to depend upon properties of the host
cluster, it is imperative that samples are drawn from large surveys
that span as wide a range of cluster properties as possible to probe
that dependence, and that the samples are sufficiently large to
average over outliers.  Both ABK and BCM have relatively small samples
at $z>0.5$ ($\sim$a dozen each) that are biased with respect to
cluster L$_x$ (ABK toward low-L$_x$ systems and BCM toward high-L$_x$
systems). Obviously, enlarging the samples of these high redshift
clusters has been difficult and any potential advance should be
exploited.  We examine the evolutionary history of BCGs using a new
sample of high redshift clusters drawn primarily from the LCDCS
(Gonzalez \etal 2001a). 

The LCDCS is a drift scan survey, conducted by the authors at the Las
Campanas Observatory in 1995, that covers 130 deg$^{2}$ of the
southern sky and yielded over 1000 cluster candidates.  Rather than
selecting overdensities of resolved galaxies, our cluster-finding
technique relies upon the assumption that the total light from a
distant cluster is dominated by the contribution from unresolved
sources.  In an intrinsically uniform image, the integrated diffuse
cluster light will manifest itself as a low surface brightness
enhancement.  Not only does this catalog result in a factor of 5
increase in the number of cataloged clusters at these redshifts, but
because our cluster identification criteria differs from those
utilized in previous surveys, this catalog provides an independent,
well-defined sample with which to compare the results from more
traditional surveys. Details of the survey and the catalog of clusters
can be found in Gonzalez \etal (2001a).

Because extensive follow-up observations are prohibitively
time-intensive, the BCGs we examine here are a small subset of the
final catalog. We use 54 clusters from the LCDCS and 9 northern
clusters from a smaller Palomar 5m survey (for which the data was originally taken as part of
the Palomar Transit Grism Survey for high redshift quasars; Schneider
\etal 1994 and analyzed in Dalcanton 1995) which we originally utilized to test the low surface
brightness technique. The combined sample consists of 63 clusters, 17
of which are spectroscopically confirmed and 46 of which have
photometrically estimated redshifts (Nelson \etal 2001a). We obtain
deep follow-up imaging in $V$, $I$, and \Kp, although not every BCG is
imaged in every band. In \S2 we describe the data. We include a brief
description of the survey, cluster candidate selection and
classification in \S2.1. In \S2.2 we describe follow-up observations
and photometry. Next, we give a summary of our photometric redshift
estimators and BCG selection in \S2.3 and \S2.4, respectively. In
\S2.5 we describe the technique used to estimate L$_x$. In \S3 we
examine the evolution of BCGs in both luminosity and color.  Finally,
we summarize our results and conclusions in \S4.

\section{The Data}

Our data originate from a variety of telescopes and instruments. We
identify the candidate galaxy clusters using drift scans and
techniques described briefly below for context, but in full detail by
Gonzalez \etal (2001a). The cluster sample and observations used here
stem from deep optical and infrared follow-up imaging of a small
subset of the full catalog which was obtained to aid in the
classification of candidates and to develop photometric redshift
indicators. We will briefly review the classification criteria based
upon the deep imaging and our photometric redshift estimators. Nelson
\etal (2001a) contains a full description.

\subsection{Cluster Sample}

We draw our clusters from two optical drift scan surveys.  The first
is a $\sim17.5$ deg$^{2}$ Palomar 5m survey using the 4-shooter camera
(Gunn \etal 1987) and the F555W ``wide $V$'' filter (Dalcanton \etal 1995, 1997).  Dalcanton \etal (1995, 1997) made use of existing
scans (originally taken as part of the Palomar Transit Grism Survey
for high redshift quasars; Schneider \etal 1994) to verify the low
surface brightness (LSB) technique for detecting LSB objects to the
required central surface brightness.  This survey identified more than
50 northern hemisphere cluster candidates.  The second is a 130
deg$^{2}$ survey of the southern sky conducted at the Las Campanas
Observatory 1m telescope over 10 nights by the authors using the Great
Circle Camera (Zaritsky, Shectman, \& Bredthauer 1996) and a wide
filter, denoted as $W$, that covers wavelengths from 4500 \AA\ to 7500
\AA. This survey was specifically designed to study LSB galaxies and
high redshift galaxy clusters. We stagger scans in declination by half
of the field-of-view (24 arcmin) so that each section in the survey is
observed twice, each time with a different region of the detector. The
sidereal rate limits exposure times to $\sim$97 sec per scan at the
declination of the survey ($-10^\circ$ to $-12^\circ$), so that the
net exposure time is $\sim$195 sec at any location. The pixel scale is
0.7 arcsec/pixel with a typical seeing of about 1.25 arcsec. The full
region surveyed measures $\sim 85^\circ \times 1.5 ^\circ$ and is
embedded within the area covered by the LCRS (Shectman \etal 1996).
The result of iterative flatfielding, masking, filtering, and object
detection is a catalog of $\sim$ 1000 cluster candidates (Gonzalez
\etal 2001a provide a full description of the method and catalog) with
an estimated 30\% contamination rate.

From the drift scans, we select a set of candidates for which to
obtain deep follow-up imaging in $V$, $I$, and \Kp. These deeper
images allow us to classify detections and train the cluster-finding
algorithm. The cluster candidates presented in this work were
confirmed as clusters based upon the following criteria: 1) the
appearance of the background-subtracted $I-$band luminosity function,
in particular a well populated and defined bright end of the
luminosity function; 2) the presence of a prominent red envelope in
the $V-I$ CM diagrams, and 3) a well-defined concentration in the
surface density of galaxies, in particular a high degree of spatial
clustering of red galaxies.  A cluster candidate that strongly
satisfies any of these three criteria is considered a viable
cluster. We do not require successful candidates to satisfy all three
criteria for two reasons. First, we do not have the necessary data to
evaluate every cluster equivalently. Second, we have confirmed
clusters that do not strongly satisfy all three criteria (for example,
high redshift clusters generally have less prominent red envelopes due
to the combination of cosmological shifting of light out of the
$V$-band and our modest exposure times; Nelson \etal 2001a).  We
acknowledge that our classification procedure is subjective, but
simulations demonstrate that random fields and non-cluster detections
(LSBs and spurious detections) do not yield well-defined luminosity
functions nor prominent red envelopes (Nelson \etal
2001a). Nevertheless, we will discuss the effect of significant
contamination on our results in \S3.1.1.

Because the original goal of the
follow-up imaging was to test our cluster selection algorithms, the
cluster subsample is heterogeneous and thus includes both some of our
best and some of our worst candidates.
In addition, since our selection procedure evolved during this
testing process, some of the confirmed candidates presented here are
not in the final, statistical cluster catalog presented by Gonzalez
\etal (2001a).  We expect no particular biases in BCG properties from
this procedure, except those that correlate with cluster properties.
The potentially influential cluster properties include concentration
and cluster mass, which factor into the cluster selection and which
are present at some level in all cluster surveys.  These biases are
discussed in detail in \S2.4.

\subsection{Optical and Infrared Photometry}

We briefly review the data acquisition and reduction process. Readers
interested in a more detailed description are referred to Nelson \etal
(2001a). We image our clusters in the Johnson $V$ and Cousins $I$-band
because these bands straddle the Ca II H \& K break for galaxies at
redshifts $0.3 \leq z \leq 0.9$, and \Kp because it probes the older,
quiescent stellar populations.  Because the majority of our photometry
was completed before the cluster finding algorithm was fine-tuned
(indeed it was necessary to help train the algorithm), we imaged some
candidates only in $I$, to provide more images of potential clusters
at the expense of color information. The data were collected at a
variety of telescopes from 1996 to 1998.  Northern clusters were
generally imaged at the Palomar 1.5 m telescope and the Lick 3 m
telescope, while southern clusters were generally imaged at the Las
Campanas Observatory using the 1 m and 2.5 m telescopes. Both the
optical and infrared data were reduced following standard
procedures (see Nelson \etal 2001a for details). We calibrated the optical data using Landolt's standard
fields (Landolt 1983, 1992).  The \Kp-band data was calibrated using
the HST faint near-IR standards (Persson \etal 1998).

Table 1 contains information regarding the observations of each
cluster.  The first column lists the cluster designation. RA and DEC
(2000.0) are in the next column, where the center of the cluster is
taken to coincide with the centroid of the low surface brightness
fluctuation.  The last six columns contain the exposure time and
seeing for each of the three photometric bands, $V$, $I$, and \Kp.  In
summary, 97$\%$ of our clusters are imaged in $I$, $54\%$ in $V$,
$35\%$ have infrared \Kp data, and $29\%$ have spectroscopy. Some
clusters were imaged twice in various photometric bands and are listed
twice in Table 1.  If one data set is clearly of superior quality than
the other, we use the superior data.  Otherwise we average the
photometric results.

The correlation between the luminosity of the BCG measured within the
fixed radius $r_{m}$, referred to as the metric aperture or L$_m$, and
the logarithmic slope of the BCG curve of growth, $\alpha \equiv[$d
log(L$_{m})/$d log$(r)]|_{r=r_{m}}$ (Hoessel 1980; Lauer \& Postman
1994; Postman \& Lauer 1995, hereafter PL95; Hudson \& Ebeling 1997;
CM) is often used to place BCG luminosities on a standard-candle
scale.  Such an adjustment typically reduces the scatter in the
distance-magnitude relation by $\sim0.1-0.2$ mag.  Consequently, BCG
photometry is predominantly measured using a fixed aperture of
projected physical radius ($\sim$10 kpc locally, and $\sim$25 kpc at
high redshift). However, we choose neither to measure our BCG
photometry with fixed metric apertures, nor to correct their
luminosities using the L$_{m}-\alpha$ relation, for two
reasons. First, seeing corrections that must be applied to fixed
aperture magnitudes are too uncertain (especially at high redshift,
because the seeing varied dramatically from run to run,
0.8$^{\prime\prime}-3^{\prime\prime}$ in the optical and
0.6$^{\prime\prime}-2^{\prime\prime}$ in the infrared). Second, as we
describe in detail below, the majority of clusters in our sample do
not have spectroscopic redshifts. Instead we rely upon photometric
redshift estimators whose associated errors ($\sigma_{phot}\sim0.06$)
are comparable to the L$_{m}-\alpha$ correction. Applying the
L$_{m}-\alpha$ relation to BCG photometry typically changes M$_{BCG}$
by $\sim$0.5 mag (see Figure 1 in PL95; Figure 10 in CM). 
For the redshift range in which the bulk of our clusters lie,
the uncertainty in $z_{phot}$ corresponds to $\Delta M_{bcg} \sim
$0.2-0.4.  Despite the potential increase in scatter by not using the
L$_m - \alpha$ relationship, we argue that such precision is lost in
comparison to simulations (or stellar evolution models) that do not
model evolution in the structural properties of galaxies. Our sample
is useful for global tests, but has high photometric scatter that
prevents case-by-case comparisons.

Instead of an aperture magnitude, we measure magnitudes using
SExtractor's (Bertin \& Arnouts 1996) ``best'' total magnitudes.  For
uncrowded objects, this magnitude is the automatic aperture magnitude,
which has been demonstrated to miss less than 6\% of the light from a
galaxy (Bertin \& Arnouts 1996). The corrected isophotal magnitude,
which misses less than 15\% of the galaxy's light, but is more robust,
is used for crowded objects (Bertin \& Arnouts 1996). These techniques
are not necessarily correct for galaxies with large extended halos,
such as those BCGs may have, but the low surface brightness halo is
missed at all redshifts (see Gonzalez {\it et al.} (2001a) for a
demonstration that $\sim$ 30\% of the light from a nearby cD is missed
by standard techniques).  Aside from this last uncertainty, 
the internal uncertainties as calculated by SExtractor using the detector
read noise and gain, and the photon statistics of sky and object, dominate.
However, we also include the uncertainty from the
photometric solutions in determining the uncertainty in the final
magnitudes. Galaxy magnitudes are corrected for extinction using the
dust IR emission maps of Schlegel \etal (1998), but are not
K-corrected.

\subsection{Spectroscopic and Photometric Redshifts}

To study the evolution of the BCGs, we must know their relative
distances.  We either obtain redshifts directly from spectroscopy of a
number of potential cluster members, or indirectly from photometry of
cluster members.  Because the final cluster catalog contains
$\sim$1000 candidates, it is not feasible to obtain spectroscopy for
all our potential clusters, nor even for the 65 clusters discussed
here.

We obtained spectra of 19 of our clusters using the Low-Resolution
Imaging Spectrograph (Oke \etal 1995) on the Keck I telescope in 1995
Dec 20-21, 1997 March 14-15, 1998 April 4-5, and 1999 March 20-21.
Due to weather and technical problems $\sim$ 40\% of the time was
usable.  The deep imaging necessary to construct multi-object masks
was generally not available so we used a single long slit with a 600
line mm$^{-1}$ grating.  The slit was aligned, using the guider image,
to include as many individual galaxies as possible, generally 2 to 3
with m$_{R} < 22$, and to lie across the position of the LSB feature
detected in the scans. Typical spectroscopic exposure times were 30
minutes to one hour.  Details of the reduction and analysis of the
spectroscopy can be found in Zaritsky \etal (1997) and Nelson \etal
(2001a).

For clusters without spectroscopy, we estimate the redshifts
photometrically using two different techniques.  First we use the
$I-$band luminosity function measured from the $I$-band follow-up
observations to obtain $z_{lum}$.  Second, we use the location of the
``red envelope'' in the cluster color-magnitude (CM) diagram measured
using the $V$- and $I$-band follow-up observations to obtain
$z_{col}$.  For details of the procedure see Nelson \etal (2001a).
Each technique has its strengths and weaknesses. The more practical
$z_{lum}$ is more efficient than $z_{col}$ because it requires deep
follow-up imaging in only one band.  However, the cluster galaxy
catalog must be corrected for interloper contamination by
statistically subtracting the background galaxies, which introduces a
significant source of uncertainty. On the other hand, $z_{col}$ does
not require background subtraction because the red ellipticals in the
cluster are typically the reddest objects in the field, but the
technique is less efficient because it requires follow-up observations
in two separate filters.  We empirically calibrate both methods using
a subset of our clusters with spectroscopic redshifts and the 19
clusters taken from the literature (A93; Smail \etal 1997; Stanford
\etal 1998; see Nelson \etal 2001a for details).

Using clusters with $z_{spec}$ we find that $z_{lum} - z_{spec}$ has
an rms scatter about zero of 0.06, which we adopt as our error in
$z_{lum}$.  Similarly, we calculate $z_{col} - z_{spec}$ and find the
rms scatter about zero is 0.07, and adopt this value as our error in
$z_{col}$. These values of the rms underestimate the redshift
uncertainties (because the spectroscopic clusters are perhaps richer
than the average candidate), but our conclusions do not depend on
having precise redshifts or exact redshift uncertainties. All clusters
with $z_{spec}$ were used for these estimates. For clusters without
$z_{spec}$, we rely on the two photometric redshift estimations.  For
our full redshift range, if both $z_{lum}$ and $z_{col}$ exist, we
average the two determinations.  We disregard any clusters with
$z_{lum}>0.7$ that do not also have $z_{col}$ because $z_{lum}$
becomes unreliable at high redshift; our observations are generally
too shallow to contain a sufficient number of galaxies to produce a
well populated luminosity function. This last cut removes nine BCGs
from the final analysis.  The redshift distribution of our BCGs with
spectroscopic redshifts (\textit{filled histogram}) and the full
sample (\textit{open histogram}) is shown in Figure 1.

\subsection{Estimating Cluster L$_x$} 

As has been recently discussed by BCM and will become more evident in
\S3, environment, as classified on the basis of L$_x$, is related to
the properties of BCGs.  Although x-ray observations do not yet exist
for our clusters, we found that the integrated light of the cluster
can be used as a proxy for x-ray luminosity (Gonzalez \etal 2001b).
Gonzalez \etal obtained drift scan data for 17 known clusters with
x-ray observations (only one of which lies within the original survey
area).  To directly compare to the results from previous high redshift
BCG studies (CM; BCM), we calculate L$_{x}$ for the EMSS passband
(0.3$-$3.5 keV) assuming H$_{0} = $50 km s$^{-1}$ Mpc$^{-1}$,
$\Omega_{m} = 1$, and $\Omega_{\Lambda} = 0$. Non-EMSS clusters are
converted to the 0.3$-$3.5 keV passband using XSPEC assuming a Mekal
model (Mewe, Gronenschild and van den Oord 1985; Mewe, Lemen and van
den Oord 1986; Kaastra 1992) with a neutral hydrogen column density,
n$_{H} = 5 \times 10^{20}$ atoms cm$^{-2}$, and a metal abundance,
$[$Fe/H$]$ = 0.3. For clusters without published x-ray temperatures,
we assume T$ = $5 keV. Because the majority of the non-EMSS clusters
were observed by ROSAT (0.2$-$2.5 keV passband), our results are only
weakly dependent upon these assumptions.

In Figure 2 we plot L$_{x}$ vs $\Sigma(1 + z)^{5}$, where $\Sigma$ is
the peak surface brightness of the detection fluctuation in counts
s$^{-1}$ arcsec$^{-2}$. For L$_{x} \gtrsim 2 \times 10^{44}$ ergs
s$^{-1}$ the correlation with $\Sigma(1 + z)^{5}$ is good. Below this
threshold $\Sigma(1 + z)^{5}$ systematically underestimates L$_{x}$ and
the scatter increases significantly. The threshold is remarkably
similar to that which divides the BCG sample of CM into a high-L$_{x}$
homogeneous population and a low-L$_{x}$ heterogeneous population,
suggesting that the breakdown in our relation is perhaps physically
motivated.  We adopt L$_{x} = 2 \times 10^{44}$ ergs s$^{-1}$ as the
threshold between a high-L$_{x}$ and low-L$_{x}$ sample, which
corresponds to $\Sigma(1 + z)^{5} \sim 0.100$ counts s$^{-1}$
arcsec$^{2}$.

Using the empirically calibrated L$_{x}$-$\Sigma(1 + z)^{5}$ relation
in conjunction with the surface brightness limit of the cluster
catalog, we estimate the limiting x-ray luminosity of the entire LCDCS
as a function of redshift (Gonzalez \etal 2001a). At low
redshift the LCDCS is sensitive down to the level of poor groups,
while at $z \gtrsim 0.6$ we are primarily detecting high-L$_{x}$
clusters only. In Figure 3 we plot $\Sigma$ as a function of redshift
and search for a similar trend in the cluster subsample examined in
this work. For $z \lesssim$ 0.6, we detect clusters with a wide range
in L$_x$, while above this redshift we become increasingly less
sensitive to the lower mass clusters.  Consequently, any correlations
between BCG properties and cluster mass will result in apparent trends
with redshift if we do not correct for this selection bias.

\subsection{BCG selection}

Selecting the galaxy that is truly the BCG requires careful
consideration of several factors: passband used for selection, search
radius, and contamination. Even at low redshift, where extensive
spectroscopy is available, BCG selection is problematic. For example,
PL95, who studied 119 BCGs with $z \le$ 0.05 (which included a subset of
Hoessel, Gunn, \& Thuan's (1980) sample) found that because their selection
criteria differed from that of Hoessel, Gunn, \& Thuan (1980), they
selected a different galaxy as the BCG in seven out of 34 clusters.

Our first consideration is which passband to use for
selection. Although bright cluster ellipticals have remarkably
homogeneous colors (A93; Stanford \etal 1995; Lubin 1996; Stanford
\etal 1998;
Kodama \etal 1998; Nelson \etal 2001a), even a small spread in colors
can result in a different choice of BCG depending on the bandpass. Of
the three bandpasses available to us, \Kp is the optimal because it is
most sensitive to the older stellar populations throughout the
redshift interval of our sample. However, because of the small
field-of-view of IR arrays (corresponding to $150-250 h^{-1}$ kpc at
the cluster redshifts), \Kp selection will miss BCGs that are
significantly displaced from the cluster center.  Instead, we choose
to use the $I$-band for selection because we have large field-of-view
images in the $I$-band of our entire sample.  The $I$-band lies
redward of the Ca II H \& K break for the majority of clusters in our
sample ($z \lesssim 0.8$) and consequently is not overly sensitive to
recent star formation.

Our second consideration brings us to the greatest difficulty in
selecting high redshift BCGs: foreground contamination. Unlike for
nearby clusters, extensive spectroscopy necessary to confirm cluster
membership is unavailable for the majority of clusters at high redshift.  Furthermore,
PL95 found that although many BCGs are projected close to the cluster
center ($\langle r_{BCG} \rangle < 70 h^{-1}$ kpc), a few lie at large
projected distances ($r_{BCG}\sim$1 Mpc) where contamination is
high. Choosing the optimal search radius, so as to include as many
BCGs as possible while minimizing contamination, becomes a crucial
issue.  Color information is extremely useful in mitigating
contamination because field contaminants are lower redshift galaxies
that are bluer than the high-redshift ellipticals.  We have colors
($V-I$ and/or $I-$\Kp) for 60\% of our clusters. To determine the
optimal search radius, we first identify the BCG in clusters for which
color information is available. We then use the radial distribution of
these color-selected BCGs to determine the optimal search radius with
which to maximize our chance of finding the proper BCG for clusters
without color information (i.e. the radius at which we minimize the
combined impact of foreground contaminants and failure to include the
BCG).

For the first step in this sequence, color-aided BCG selection, we set
a large search radius because the selection is less susceptible to
interlopers.  PL95 found that 90\% of their 119 local BCGs ($z \le
0.05$) lie within a projected separation of $350 h^{-1}$ kpc.
Therefore, for our clusters with either $V-I$ and/or $I-$\Kp colors
available, we consider all galaxies within $350 h^{-1}$ kpc of the
cluster center, where the cluster center is defined to be the centroid
of the low surface brightness detection on the drift scan images.  The
sole modifier to this criteria is that when we are using the \Kp
images, our small field-of-view limits our search to the central
$150-250 h^{-1}$ kpc of the cluster.  Because bright cluster
ellipticals occupy a very narrow locus in the cluster color-magnitude
diagram (cf. SED98), we exclude BCG candidates whose colors are 0.4
mag bluer than the location of the red envelope in either $V-I$ or
$I-$\Kp.  The location of the red envelope in color space is defined to
be the position of maximum change in the number of galaxies as a
function of color (see Nelson \etal 2001a for details).  Of the
remaining galaxies, we choose the galaxy with the brightest total
magnitude in the $I$-band as the BCG.

We examine both the distribution of $r_{BCG}$ and the contamination
rates as a function of selection radius for color-selected BCGs,
$r^{col}_{sel}$, to find the optimal search radius for those clusters
without color information, $r^{non-col}_{sel}$.  Figure 4 shows the
distribution of $r_{BCG}$ for BCGs selected by color within $350
h^{-1}$ kpc of the cluster center.  The open histogram shows the radial
distribution for all color-selected BCGs, while the filled histogram is for
color-selected BCGs with $z_{spec}$.  Although there is a high
concentration ($\sim$40\%) of BCGs very close to the cluster center
($r_{BCG} \leq 50 h^{-1}$ kpc), there is also a significant number of
BCGs ($\sim$40\%) at larger projected separations ($r_{BCG}>150
h^{-1}$ kpc).  Next, we vary the search radius and estimate
contamination rates, on a per cluster basis.  First, we identify the
``true'' cluster BCG using the color-selection criteria outlined above
(including $r^{col}_{sel} = 350 h^{-1}$ kpc). Then, we choose 10
random centers on the image that lie entirely outside of the cluster
radius ($350 h^{-1} $kpc$ + r^{non-col}_{sel}$) and select the
brightest $I$-band galaxy that lies within $r < r^{non-col}_{sel}$
\textit{without} using color information. Finally, we define the
contamination rate for this cluster as the fraction of times that a
galaxy within the search radius, $r^{non-col}_{sel}$, is brighter than
the ``true'' (i.e.  the original, color-selected ) BCG.  We repeat
this entire process for every cluster using various search radii.
Figure~5 shows the distribution of contamination rates for four values
of $r^{non-col}_{sel}$. For very small search radius,
$r^{non-col}_{sel} = 50 h^{-1}$ kpc, the contamination rates are
extremely low, $\sim$80\% of the clusters have effectively no
contamination and there are no clusters with a contamination rate
greater than 20\%.  By the time $r^{non-col}_{sel}$ reaches $200
h^{-1}$ kpc, the contamination rates are quite high, 55\% of the
clusters have contamination rates $>$30\%. We also calculate $\langle
\Delta {\rm m} \rangle = {\rm m}_{BCG} - {\rm m}_{cont}$ and find that $\langle
\Delta$m$ \rangle \sim 1$ mag in $V$ and $I$.

Because there is a significant difference between the magnitude of the
true BCG and that of a foreground contaminant, the optimal search
radius for non-color-selected BCGs must be small enough to keep
contamination relatively low, yet large enough to include a
significant fraction of BCGs.  While the contamination rate for
$r^{non-col}_{sel} = 50 h^{-1}$ kpc is attractive, such a sample would
contain only $\sim$40\% of the BCGs.  Instead, we choose to use
$r^{non-col}_{sel} = 150 h^{-1}$ kpc, which contains a significant
fraction of BCGs ($\sim$60\%) yet still has relatively low
contamination rates (70\% of the clusters have a contamination rate
$\leq$20\%).  When we miss the BCG, and instead recover the second or
third rank galaxy, the difference in magnitude is on average 0.5 mag
(cf. \S 3.1).

Although the majority of BCGs are either elliptical or cD galaxies (cf.
Graham \etal 1996), this is not necessarily universal. Because we do
not have morphological information for cluster galaxies, our BCGs are not
restricted to be of a certain type. However, color selection of BCGs could
bias our results if some BCGs are bluer than our color criteria. An
inspection of the cluster color-magnitude diagrams shows that for 
50\% of the clusters, the red-selected BCG is simply the brightest
galaxy within 350$h^{-1}$ kpc, regardless of color.
Therefore, color does not play a role in the selection of 70\%
of the BCGs in our sample (including the 40\% of our sample for which
we do not have color information). For the remaining 50\% of the 
color-selected BCGs that are not the brightest galaxy within 350$h^{-1}$
kpc, the brighter galaxy is bluer by $>$ 1 mag in $V-I$
%, and therefore a likely foreground galaxy, 
in 65\% of the
cases. Consequently, our sample is not strongly biased toward
red BCGs.

In Figure~6 we present $I$-band postage stamp images of the BCGs used
in this study (except for 0944$+$4732 whose image did not reproduce
well). The images are $1.5^{\prime} \times 1.5^{\prime}$ and
are centered on the BCG. Table 2 lists the properties of the
BCGs. Column 1 gives the cluster name. The BCG redshift and corrected
cluster central surface brightness, $\Sigma(1 + z)^{5}$, are listed in
Columns 2 and 3. The next three columns give the magnitude of the BCG
in $V$, $I$, and \Kp. The final two columns contain the color of the
BCG in $V-I$ and $I-$\Kp. In summary, 61\% of our sample is imaged in
$V$, 96\% in $I$, and 35\% in \Kp. We have color information (in
either $V-I$ or $I-$\Kp) for 63\% of the BCGs. Finally, we have
L$_{x}$ estimates of 61\% of the BCGs host clusters.

\section{Results}

According to hierarchical clustering models, the characteristic scale
upon which structure forms increases with time. Consequently, low
redshift clusters of a given mass have experienced a different
evolutionary history than equivalent mass clusters at high
redshift. Galaxy evolution studies of cluster and field galaxies at $z
\lesssim 1$ have supported this idea by providing evidence that
environment is linked to the properties of galaxies (e.g. ABK; CM;
BCM). Here we examine the luminosities and colors of BCGs from $z \sim
$ 0.3-1 and search for differences in their evolutionary histories as
a function of environment.

\subsection{BCG Luminosity Evolution}

We search for the influence of environment on evolution by examining
$V$, $I$, and \Kp Hubble diagrams (Figure 7)
and differentiating between BCGs from high-L$_{x}$ clusters,
low-L$_{x}$ clusters, and clusters for which we do not have L$_{x}$
estimates\footnote{In order to estimate L$_x$, we must be able to
accurately measure the surface brightness of the cluster
detection. However, we cannot do this for about one-third of the
cluster candidates identified in our earliest analysis, a subset of
which are used in this work, because those
objects are near bright stars. Subsequent catalogs (including the
final catalog presented by Gonzalez \etal 2001a) are drawn from
analyses using larger masked regions.}.  Similar to CM and BCM, we
adopt L$_{x} = 2 \times 10^{44}$ ergs s$^{-1}$ as the threshold
between a high-L$_{x}$ and low-L$_{x}$ sample. The errors
in the photometry are negligible, but the uncertainty in the
photometric redshifts is not. An error in $z_{phot}$ of 0.07 corresponds
to $\Delta m_I \sim$ 0.3 mag at $z \sim 0.6$.

We compare the BCG luminosities to spectral synthesis models using the
GISSEL96 model (Bruzual \& Charlot 1993) for an elliptical galaxy
experiencing either no-evolution or passive evolution. For all models,
we assume an initial 10$^{7}$ yr burst of star formation and a
Salpeter initial mass functions for masses between 0.1$\msun$ and
100$\msun$. We determine the magnitude zero-points for the models by
using 64 of PL95's local BCGs whose host clusters have x-ray
luminosities from ROSAT.  Only two of these clusters have L$_{x} \geq
2 \times 10^{44}$ erg s$^{-1}$ in the EMSS passband and consequently
this is essentially a low-L$_{x}$ normalization. For each BCG, PL95
provide photometry in a series of apertures of increasing angular
size. Because we use total magnitudes, rather than fixed metric
magnitudes, we use the largest available aperture from PL95. We
convert their $R_{c}$ photometry to $V$, $I$, and \Kp using the color
predictions of the GISSEL96 model (Bruzual \& Charlot 1993) for a
passively evolving elliptical galaxy that formed at $z_{form} =
10$. Finally, we use the locus of the converted low redshift BCGs to
normalize the models.

First, we investigate whether the predictions from a single
evolutionary model fit the observed magnitudes of \textit{all} the
BCGs, independent of the L$_x$ of their host cluster. Because at any
given redshift there is considerable observational scatter in the
magnitudes, we also present a binned version of the full data set in
redshift intervals of $\Delta z = $0.1 in $V$ and $I$ (Figure 7
\textit{upper right panel and center right panel}) and $\Delta z = $0.2 in \Kp (Figure 7,
\textit{lower right panel}). We adopt $\sigma_{mean}$ of the binned
data as our errors and find that they are typically smaller than the size of the points and are
therefore omitted. In the \Kp-band, the mean BCG luminosities are most consistent
with the no-evolution prediction, but systematically fainter than the
model predictions.  To determine whether this offset reflects some
problem with our data, we compare to the published $K$-band photometry
of A93 for 19 other BCGs with $z = 0.02-0.92$ (Figure 7, \textit{lower right panel}). The published photometry was measured inside a metric
aperture of 50 kpc diameter assuming $\Omega_{m} = 1, \Omega_\Lambda=0$ 
and H$_{0} = $50 km
s$^{-1}$ Mpc$^{-1}$ and corrected for galactic extinction using the
reddening maps of Schlegel \etal (1998), but were not K-corrected.
Whereas our BCGs show higher scatter than those of A93, on average our
results are consistent.  A small change in zero-point would bring the
no-evolution model in agreement with both our data and that from A93.
Only two of our \Kp BCGs are in high-L$_x$ clusters and the A93 sample
also consists principally of low-L$_x$ clusters. Therefore, we
conclude that the \Kp luminosities of BCGs from low-L$_{x}$ clusters
are sufficiently homogeneous to be modeled as an apparently
non-evolving population (confirming A93 and ABK's results, but
restricting those results to the low-L$_x$ clusters, in accordance
with the results of BCM).

We have a different situation for BCGs observed in $V$ and $I$. The
BCGs have mean luminosities that are consistent with the no-evolution
predictions for $z \lesssim 0.6$ (Figure 7, \textit{upper right
panel and center right panel}) but are systematically brighter than the no-evolution
prediction and consistent with the passive evolution of an elliptical
galaxy that formed at $z_{form} = 5-10$ at $z \gtrsim 0.6$ . For these
BCGs, a single evolutionary model is insufficient to explain the
observed magnitudes across the entire range of redshift. The upper
panels of these diagrams illustrate that, unlike for the \Kp sample,
these samples contain significant fractions of both high and low-L$_x$
clusters. Because our cluster sample is biased toward more massive
clusters at $z \gtrsim 0.6$ (cf. Figure 3), it is likely that the high
redshift clusters without L$_{x}$ estimates are also high-L$_{x}$
systems. If this supposition is true, it suggests that BCGs from
high-L$_{x}$ clusters follow a different evolutionary scenario than
those from low-L$_{x}$ clusters. We suggest that the
inclusion of both populations is responsible for the apparently
erratic Hubble diagrams.

The differences between the two populations can be examined in a
variety of ways.  First, we examine the deviations from a single
evolutionary model in the Hubble diagrams more closely. The BCG
$V$ and $I$ magnitude residuals, $\Delta m = m_{BCG} - m_{model}$, for a passive
evolution model with $z_{form} = 5$ is shown versus redshift in
Figures 8 and 9 for low-L$_{x}$ clusters
(\textit{left panel}) and high-L$_{x}$ clusters (\textit{right
panel}). We omit low-z ($z < 0.6$) clusters for which we could not
measure $\Sigma$, but we do include clusters with $z \geq 0.6$ for
which we could not measure $\Sigma$ (\textit{filled triangles})
because at high redshift we expect to only detect very massive
clusters (cf. Figure 3; Gonzalez \etal 2001b). The BCG magnitudes
(except for the $V$-band high-L$_{x}$ sample) show significant
scatter, $\sim 0.40$ mag. Because errors in the photometry are
negligible, this scatter is most likely due to the uncertainties in the
photometric redshifts (at $z \sim 0.6$, $\Delta z = 0.06$ corresponds
to $\Delta m_{I} \sim 0.3$ mag). The filled boxes
denote the mean BCG magnitude residuals in $V$
and $I$ with error bars that are $\sigma_{mean}$. One high-L$_{x}$ cluster (1024$-$1239) has very discrepant magnitude
residuals ($>$8$\sigma$) in $V$ and $I$ and therefore is omitted from
the determination of the mean magnitude residuals for the high-L$_{x}$
sample, under the assumption that it represents an incorrect
identification of the BCG or a spurious detection. In the $V$-band, $\langle \Delta$m$_{V}^{{\rm
low-L}_{x}}\rangle=0.42\pm0.12$ and $\langle \Delta$m$_{V}^{{\rm
high-L}_{x}}\rangle=0.07\pm0.10$ for low-L$_{x}$ and high-L$_{x}$
clusters respectively, while $\langle \Delta$m$_{I}^{{\rm
low-L}_{x}}\rangle=0.58\pm0.09$ and $\langle \Delta$m$_{I}^{{\rm
high-L}_{x}}\rangle=0.31\pm0.11$ in the $I$-band.  We conclude that
BCGs from high-L$_{x}$ clusters are on average brighter than those
from low-L$_{x}$ clusters and more closely follow the predictions of a
passive evolution model with $z_{form} = 5-10$.

Second, we examine the BCG magnitude residuals from the model
predictions as a function of $\Sigma(1 + z)^{5}$. In Figure 10, we
plot $\Delta m = m_{BCG} - m_{model}$ vs. $\Sigma(1 + z)^{5}$ in $V$
(\textit{left panel}) and $I$ (\textit{right panel}) for passive
evolution with $z_{form} = 5$.  Filled triangles represent high-z ($z
\geq 0.6$) BCGs without measured values of $\Sigma$. Because our
cluster sample is biased toward very massive systems at $z \gtrsim
0.6$, we consider these BCGs to be from high-L$_{x}$ clusters and
assign them lower limiting values of $\Sigma(1 + z)^{5} =
0.100$. $\Delta m_{V}$ correlates inversely with $\Sigma(1 + z)^{5}$
at the 90\% confidence level according to the Spearman rank test with
or without the upper limit points included. In $I$, the correlation is
less significant. Including the upper limits, the correlation is
significant with 90\% confidence, but excluding them the correlation
is significant at only the 85\% level. These correlations illustrate
that as $\Sigma(1 + z)^{5}$ increases, the residuals from the passive
evolution model, tend to decrease.

We summarize the observational situation as follows: 1) the BCG sample
includes clusters of both high and low L$_x$, where the dividing line
in L$_x$ is set, both from previous studies and the behavior we
observe between L$_x$ and $\Sigma$, at L$_x = 2\times 10^{44}$ ergs
sec$^{-1}$; 2) \Kp observations of BCGs (both our data and that of
A93) are consistent with no-evolution models, but are based on BCGs
that are almost exclusively in low-L$_x$ clusters; 3) the $V$ and $I$
BCG Hubble diagrams, which include BCGs from both high and low-L$_x$
clusters, cannot be fit with a single class of evolutionary model; and
4) after dividing the sample according to host cluster L$_x$ we find
that the BCGs from low-L$_x$ clusters fit the no-evolution model in
both $V$ and $I$ (in agreement with the results from our \Kp
observations), and that the BCGs in the high-L$_x$ clusters are better
fit by a passive evolution model, particularly in $V$ where recent star formation should be more
detectable. To address whether these
results are driven by physical effects, we next focus on potential
problems with our data and approach.

\subsubsection{Potential Pitfalls}

We observe that the luminosities of high redshift BCGs ($z \gtrsim
0.6$) are, on average, brighter than the no-evolution extrapolation of
the luminosities of their low redshift counterparts. Because our sample
is biased toward more massive systems at high redshift, we interpret
this as evidence that environment may influence the evolution of
BCGs. However, there are other factors that could explain the observed
``brightening" of BCG luminosities at $z \gtrsim 0.6$.  We discuss the
possibilities below.

\paragraph{False Clusters:}

The fraction of false detections of clusters in the original survey
increases as we consider larger redshifts (cf. Gonzalez \etal 2001a)
and one might suspect that this effect could lead to various
misleading trends in the ``apparent" BCG population.  Because of our
follow-up imaging, we do not consider contamination to be as serious
for this cluster subsample as for the survey itself. First, our
redshift estimators (the luminosity function and red envelope method)
fail to converge on random (i.e. spurious) centers and on non-cluster
(low surface brightness galaxies) detections in our follow-up images
(Nelson \etal 2001a). Second, the luminosities of randomly selected
brightest field galaxies do not follow the predictions of either
passive evolution or no-evolution models for BCGs. To illustrate this
point, we use the simulations outlined in \S2.5 and select the
brightest field galaxy in an area comparable to that of each
cluster. Assuming the randomly selected galaxies lie at the cluster
redshift, the dispersion in their magnitudes is significantly larger,
$\sigma_{I} = 1.2$ mag, than that of our cluster BCGs (Figure 11,
\textit{upper panel}). Their magnitudes ($\langle$m$_{I}\rangle=19.2$
mag) are typically fainter than our BCGs, and do not correlate with
the cluster redshift (Figure 11, \textit{lower panel}). Indeed, Figure
11 demonstrates that the typical luminosity of a bright field galaxy
is fainter than or consistent with the predictions of the no-evolution
models for $z \lesssim 0.8$. Figures 8 and 9 highlight such a potential
false cluster detection. 1024$-$1239, a high-L$_{x}$ BCG at $z = 0.57$, has
magnitude residuals in $V$ and $I$ that are $>$8$\sigma$ fainter than
the mean magnitude residuals of other high-L$_{x}$ BCGs. 
From our simulations, 35\% of randomly selected brightest field
galaxies at $0.50 < z < 0.60$ have magnitudes that are $> 8\sigma$
fainter than the mean magnitude of the BCGs in the same redshift
range. Consequently,
although our sample may include a few false cluster detections, 
an increase in the fraction of false clusters with $z$ in unlikely to
cause significant brightening of our observed BCGs.

\paragraph{Biased Measurement of $\Sigma$:}

If $\Sigma$ is dominated
by the BCG halo, then there will be a
correlation between brighter BCGs and $\Sigma$. Gonzalez \etal (2000)
performed a detailed analysis of the distribution of luminous matter
in the nearby galaxy cluster Abell 1651 and found that the BCG
contributed 30\% of the total cluster light within $r = 500 h^{-1}$
kpc of the cluster center. Thus, in this instance $\Sigma$ could
be contaminated by as much as 30\% by the BCG. If the BCG halo
contributes excess light to the measurement of $\Sigma$ then we expect
$\Sigma$ to decline with increasing $r_{BCG}$. Although a slight trend
appears in Figure 12, we find no statistically significant correlation
between $\Sigma(1 + z)^{5}$ and $r_{BCG}$. Therefore, although the
question of $\Sigma$ contamination by the BCG halo remains open, we do
not find evidence in our sample of biased measurements of $\Sigma$. 

\paragraph{Poor Photo-z's:}

A systematic overestimation of the cluster photometric redshifts at $z
\gtrsim 0.6$ would cause the inferred luminosities of the BCGs to be
brighter than their actual luminosities. We believe this is not a
problem for two reasons. First, contamination systematically causes an
underestimation of the redshift because foreground galaxies tend to be
brighter than cluster galaxies (see Gonzalez \etal 2001a for an
illustration of the magnitude of the effect for photo-z's determined
from the BCGs).  Second, we find no difference in the behavior of the
magnitude residuals for BCGs in clusters with photometric and
spectroscopic redshifts.  In Figure 13 we plot the magnitude residuals
in $V$ (\textit{left panel}), $I$ (\textit{center panel}), and \Kp
(\textit{right panel}) for a passively evolving elliptical galaxy with
$z_{form} = 5$, differentiating between BCGs with $z_{spec}$
(\textit{filled circles}) and those with $z_{phot}$ (\textit{open
circles}). Recall that we disregard any clusters with $z_{lum}>0.7$
that do not also have $z_{col}$, because $z_{lum}$ becomes unreliable
at high redshift. There is no statistically significant difference
between the luminosities of BCGs in clusters with spectroscopically
and photometrically estimated redshifts.  We conclude that that errors
in our photometric redshift estimations are not causing the observed
brightening of BCGs at $z \gtrsim 0.6$.

\paragraph{Contamination:}

Another concern is that the observed brightening of BCG luminosities
in $V$ and $I$ at $z \gtrsim 0.6$ is due to increased contamination in
the BCG sample with redshift. Figure 7 shows that for $z \lesssim 0.6$
an interloping field galaxy would typically need m$_{I} \lesssim 19$
to be brighter than the expected BCG. We find that $<$5\% of all the
field galaxies in our images are brighter than the expected $I$-band
magnitude of the BCG at $z\lesssim0.6$. Consequently, we expect
insignificant contamination at low redshift. However, for
$z\gtrsim0.6$ the number of potential contaminating field galaxies
approximately doubles (13\% of our composite field galaxy population
have m$_{I}<20$ mag). 

To address this issue in a consistent manner, we use the
simulations outlined in \S2.5 to evaluate 
the contamination rate
of both non-color-selected and color-selected BCGs as a function of
redshift. We are chiefly interested in whether an increasing
contamination rate with redshift is responsible for the observed
brightening of BCGs relative to the no-evolution model prediction at
$z \gtrsim 0.6$. Therefore, we compare the magnitudes of potential
contaminants to the no-evolution model predictions rather than
directly to our data. We find that the contamination rate for
non-color-selected BCGs at low redshift ($z<0.5$) is quite low,
$\sim$10\%. However as redshift increases and we begin to sample
fainter into the luminosity function of the interloping galaxies, the
contamination rate triples, reaching $\sim$30\% at $z>0.5$. On the
other hand, the contamination rate for color-selected BCGs is quite
low and remains
relatively constant with redshift - 5\% contamination rate at $z<0.5$
and 10\% at $z>0.5$. This constancy of contamination rate arises
because bright interlopers are typically blue foreground galaxies
which are more easily differentiated as the cluster E/S0 sequence
reddens with redshift.

Are these estimated contamination rates at $z>0.6$ high enough to
explain our observed brightening? Although contamination increases to
a modest rate of 30\% at $z>0.5$ for non-color-selected BCGs, only
$\sim$20\% of BCGs at $z>0.6$ are selected in this manner. The
remaining 80\% of BCGs are selected using color information which has
a rather low contamination rate of 10\%. From our simulations, we find
that at $z>0.6$, foreground contaminants are on average 0.63 mag
brighter than non-color-selected BCGs. For color-selected BCGs,
contaminants are 0.41 mag brighter than the mean BCG
magnitude. Assuming that 30\% of non-color-selected BCGs are actually
interlopers that are 0.63 mag brighter than the BCG, while 10\% of
color-selected BCGs are foreground contaminants that are 0.41 mag
brighter than the BCG, we find that contamination causes the mean BCG
magnitude to be $\sim$0.1 mag brighter than the no-evolution model
prediction at $z=0.6-0.8$. However, we observe an actual brightening
of $\sim$0.5 mag for this redshift range in our data. Therefore, while
contamination may be responsible for some 
of the observed
brightening, it is insufficient to explain the full extent of
brightening of BCG magnitudes relative to the no-evolution model
predictions at $z>0.6$.

A final way to test whether contamination is responsible for our
observed brightening is to restrict the sample to color-selected BCGs
because they have a low contamination rate (which remains
relatively constant with redshift). The $V$ Hubble
diagram is already entirely color-selected because all of our BCGs
with $V$-band data also have $I$-band data.  The $I$ Hubble diagram,
however, does have a significant contribution from non-color-selected
BCGs.  We reproduce the $I$ Hubble diagram (Figure 14, \textit{upper
panel}) distinguishing between color-selected vs. non-color-selected
BCGs.  In the lower panel of Figure 14 we bin the color-selected
galaxies only and again find that the average BCG luminosity is
systematically brighter at $z \gtrsim 0.6$. Therefore, we conclude
that our observed brightening of BCGs at $z \gtrsim 0.6$ is not due to
increased contamination with redshift.

\paragraph{Missed BCGs:}

BCGs have been found to lie as far as $\sim$1 Mpc from the cluster
center (PL95). For those cases in which the BCG lies outside our
search radius, we will miss it and instead choose the second- (or lower-) 
ranked cluster galaxy. Locally, $\sim$90\%
of BCGs are within $350 h^{-1}$ kpc. Adopting this value
for higher redshift clusters, we expect that only $\sim$10\% of
our color-selected BCGs are, in reality, the second- (or lower-) ranked
galaxy. However, the radial distribution of our color-selected BCGs
(Figure 4) suggests that we may be missing as many as 40\% of
the BCGs for clusters that do not have color information
($r_{sel}^{non-col} = 150 h^{-1}$ kpc). How does this affect our
results? At high redshift ($z\gtrsim0.6$) our BCGs are predominantly
(80\%) color-selected. Therefore, we estimate that we select the true
BCG in at least 75\% of our high 
redshift clusters. At low redshift, the majority ($\sim$65\%) of
clusters are also selected using color information. For the modest
number of clusters that are non-color-selected, we may be missing as
many as 40\% of the BCGs. However, we find that the mean magnitude difference,
$\Delta$m$=$m$_{2nd}-$m$_{BCG}$, is $\sim$0.5 mag for both color-selected and
non-color-selected second-ranked galaxies regardless of redshift. Because we find similar values of $\Delta$m for the
color-selected and non-color-selected samples and no trend of
$\Delta$m with redshift, we conclude that if we identified the
second-ranked (or lower) galaxy rather than the BCG in some systems
this error did not lead to a detectable effect.

We conclude that systematic effects
are unlikely to account 
for the observed ``brightening"
of high redshift ($z \gtrsim 0.6$) BCGs in a sample with mixed
cluster selection that favors the selection of high-L$_x$ clusters at
higher redshifts, but warn that such problems need to be kept in mind
in all such studies (even spectroscopically-selected BCG samples 
are susceptible to some of these pitfalls).

\subsection{BCG Color Evolution}

Evolutionary constraints based on luminosities measure only one
aspect of the models. For example, a galaxy can brighten in any
particular passband by either having increased star formation or
adding, via accretion, more stars to the initial
population. Colors enable us to distinguish between these two
possibilities and so we now turn our attention to
the colors of BCGs.  Are the colors consistent with a stellar
population that formed early ($z_{form} \geq 5$) and experiences
passive evolution (as indicated by the luminosities of BCGs in
high-L$_x$ environments)? Or are they consistent with no-evolution 
and accretion (as indicated by the luminosities of the BCGs in low-L$_x$ 
environments)?

In the left panels of Figures 15 and 16 we compare the
$V-I$ and $I-$\Kp colors of BCGs to the
spectral synthesis predictions of Bruzual \& Charlot (1993) for an
elliptical galaxy experiencing either passive
or no-evolution.  There is considerable observational scatter in the colors of
these galaxies, so we bin the data in redshift intervals of
$\Delta z = 0.1$ for $V-I$ and $\Delta z = 0.15$ for $I-$\Kp (\textit{right panels}).
The behavior in both colors corresponds best with passive evolution models of an elliptical galaxy that
formed at $z_{form} \la 10$. Unlike the case of luminosity evolution,
one evolutionary model is sufficient to explain the observed colors of
all BCGs, regardless of their environment. 
It is reassuring
that the colors do not favor the non-physical model of no-evolution
for the BCGs in low-L$_x$ clusters. We conclude that environment
does not play a role in the BCG colors by observing that
there is no correlation between $V-I$ vs.
$\Sigma(1 + z)^{5}$ (Figure 17).

\section{Discussion}

Combining the constraints from luminosities and colors, the results suggest 
environment
plays a role in the mass evolutionary history of BCGs.  The 
colors of BCGs from both high-L$_{x}$ and low-L$_{x}$
clusters over the redshift range $z=0.3$-1 imply that they are predominantly comprised
of old stellar systems ($z_{form} \sim $5-10) that have undergone
little, if any, recent star formation. However, tracing the
luminosities of high-L$_{x}$ and low-L$_{x}$ BCGs as a function of
time reveals subtle, but important, differences in their
evolutionary histories. BCGs in x-ray luminous clusters,
L$_{x} \geq 2 \times 10^{44}$ ergs s $^{-1}$, follow passive
evolution model predictions typical of an old quiescent stellar
population. In addition, these BCGs have luminosities that are
systematically brighter than their low-L$_{x}$ counterparts at high
redshifts. BCGs in low-L$_{x}$ clusters do
not exhibit the characteristic dimming associated with the aging of
its stellar populations. Instead, their luminosities are fainter on
average at high redshift relative to a model of a passively evolving
population and remain relatively constant with time.

ABK interpreted the lack of passive
luminosity evolution of low-L$_{x}$ BCGs as a sign of mass accretion - BCGs accrete
mass in stars to counter the
dimming of their underlying stellar populations, thereby maintaining a
roughly constant integrated luminosity. Using $K$-band
photometry of 25 BCGs at $0 < z < 1$, ABK parameterized the luminosity
evolution by $L_{K}(z) = L_{K}(0) \times
(1+z)^{\gamma}$. Assuming $H_{0} = 50$~km~s$^{1}$~Mpc$^{-1}$, the least-squares fit to the data yielded $\gamma =
-1.2\pm0.3$ ($z_{form} = 2$) and $\gamma =
-0.7\pm0.3$ ($z_{form} = 5$) for $\Omega_{m} = 0.0, \Omega_\Lambda = 0.0$, 
while $\gamma =
-2.2\pm0.3$ ($z_{form} = 2$) and $\gamma =
-1.7\pm0.3$ ($z_{form} = 5$) for $\Omega_{m} = 1, \Omega_\Lambda = 0.0$. Provided that $K$-band light traces mass, they
conclude that BCGs increased their mass by a factor of 2 to 4 since
$z \sim 1$. In such a model, the galaxy colors would evolve passively
 if the objects accreted had a similar star formation history
as the BCG. Because ABK did not have x-ray observations of their
clusters, they assumed that their inferred mass accretion rate was
typical of all BCGs. However, BCM re-examined ABK's sample with
x-ray data and found that the majority of their clusters have
L$_{x} < 2.3 \times 10^{44}$ ergs s$^{-1}$ at $z>0.5$. To
assess the effect of cluster mass on inferred BCG mass accretions
rates, BCM studied an x-ray selected sample of 76 BCGs at
$0.05 \leq z \leq 0.83$ and parametrized the mass accretion of their
high-L$_{x}$ BCGs only. Using a parameterization of the same form as
ABK, they found that $\gamma = -0.93 \pm 0.23$ ($z_{form} = 2$) while
$\gamma = -0.38 \pm 0.22$ ($z_{form} = 5$) for 
$\Omega_m = 1.0, \Omega_\Lambda = 0.0$ and H$_{0} = 50$~km~s$^{-1}$~Mpc$^{-1}$. Consequently,
they conclude that the mass of high-L$_{x}$ BCGs has increased by a
factor of 2, at most. 

Because our data do not span a large enough baseline in redshift, we
cannot parameterize the BCG mass accretion rate as a function of
redshift for this sample. However, in Figure 18, we compare our
measured $I$-band luminosity evolution to the parametrized mass
accretion rates of ABK and BCM. The filled boxes denote the mean
magnitude residuals of our low-L$_{x}$ (\textit{left panels}) and
high-L$_{x}$ (\textit{right panels}) BCGs for a passive evolution
model with $z_{form} = 5$ (\textit{solid line}) assuming $\Omega_{m} = 0.0,
\Omega_\Lambda = 0.0$
(\textit{upper panels}) and $\Omega_{m} = 1, \Omega_\Lambda=0.0$ (\textit{lower
panels}) and H$_{0} = $ 50 km s$^{-1}$ Mpc$^{-1}$. Although the mass accretion rates predicted by ABK
(\textit{dotted line}) were derived using predominantly low-L$_{x}$
BCGs, we reproduce the relation in all four panels for reference. The
dashed line is the evolution parameterization of BCM for high-L$_{x}$
BCGs with $\Omega_{m} = 1$ only (\textit{lower right panel}). Figure 18 (\textit{left panels})
demonstrates that our low-L$_{x}$ BCGs are in excellent agreement with
the mass accretion rates of ABK, implying that BCGs from low x-ray
luminosity clusters accrete a factor of $\sim 2$ if $\Omega_{m} = 0.0$ or
$\sim 4$ if $\Omega_{m} = 1$ in mass since $z \sim 1$. 
Turning our attention to the high-L$_x$ sample (Figure 18, \textit{right panels}),
we find that the BCGs from high-L$_x$ clusters show less mass accretion than
those from low-L$_x$ clusters, which is consistent with the findings of
BCM. However, for the only case with which we can directly compare to
the results of BCM ($\Omega_m = 1, \Omega_\Lambda = 0$), our high-L$_x$ sample suggests 
a somewhat higher accretion rate than BCM (a factor of $\sim$ 2 since
$z \sim 1$ as opposed to a factor of $\sim 1.3$). In sum, our data confirm
that low-L$_x$ BCGs may have accreted a factor of 2-- 4 in mass since $z\sim 1$,
while high-L$_x$ BCGs are limited to $\lesssim 2$ increase in mass since $z\sim 1$.

Although some low-L$_{x}$ BCGs may be accreting at least
a factor of 2 in mass since $z \sim 1$, we find little evidence for any
recent star formation in the colors of BCGs. Because
the accretion is probably dominated by the infall due
to dynamical friction of massive early-type galaxies near the BCG,
large bursts of star formation are presumably unlikely.
However, even a modest amount of gas in the progenitors
could induce some star formation.  If as little as 10\% of the merger
remnant's mass is converted into stars, the galaxy's $V-I$ and $I-$\Kp
colors get bluer by as much as 2.5 magnitudes (Nelson \etal 2001a). Although
the duration of the peak bluing is short, $\lesssim$130 Myr, the galaxy
is 0.1 mag bluer than the red envelope for $\sim$1 Gyr, and we should
see that reflected in the BCG colors. Furthermore, our BCG color selection
criteria does not significantly hamper our ability to detect a galaxy
undergoing a burst of star formation. A starbursting galaxy is
$\ge$ 0.4 magnitudes bluer than the location of the red envelope for only
$\sim$260 Myr, which translates to $\sim$35\% of the duration of the
bluing period (i.e. 0.1 mag bluer than the red envelope).
From our data, we can only make statistically general claims. Even though some BCGs may appear bluer than the models, we conclude on
the basis of the average agreement with the colors from 
passive-evolution models that the
dominant form of accretion by BCGs since $z \sim 1$
consists of gas-poor, old stellar systems.

This discussion of BCG evolution suggests that, to some extent,
observational efforts at studying cluster galaxy evolution may be
self-defeating. Studies attempting to go to higher redshifts (which
naturally identify only the richest clusters) may face difficulties
because although one is selecting systems at an earlier time, these
systems may evolve faster --- thereby, countering the expected
difference between nearby and distant clusters (Kauffmann 1995a,
1995b). Despite the observations of increased merging 
in some high redshift clusters (van Dokkum \etal 1999),
galaxy evolution in the observationally accessible regime out
to $z \sim 1$ may be most dramatic in poor
clusters and rich groups (more common in the local samples).
Even at low redshift, some of these environments show
evidence for hierarchical evolution (Zabludoff \& Mulchaey 1998), and
the study of stellar populations in the giant ellipticals that result
in the dynamical collapse of such systems (cf. NGC 1132; Mulchaey \&
Zabludoff 1999) may provide a nearby example of the processes that
formed the massive ellipticals in the richest systems at $z\sim 1$.

\section{Summary} 

We investigate the influence of environment on BCG evolution using a
sample of 63 clusters at $0.3 \leq z \leq 0.9$ drawn from the Las
Campanas Distant Cluster Survey. This sample represents  a factor of 2
increase in the number of BCGs studied at $z>0.5$. Using the
peak surface brightness of the cluster detection fluctuation, $\Sigma$, as a proxy for cluster L$_{x}$
(Gonzalez \etal 2001b), we define L$_{x} = 2 \times 10^{44}$ ergs
s$^{-1}$ as the
threshold between a high-L$_{x}$ and low-L$_{x}$ sample.
We use this division to trace the luminosity and color evolution of
BCGs in $V$, $I$, and \Kp as a function of environment. Our
primary results are:

\noindent 1) We find that the luminosity evolution of BCGs from a
mixed sample of high-L$_{x}$ and low-L$_{x}$ clusters cannot be
adequately explained by a single evolutionary model.

\noindent 2) We find that the $V$, $I$, and \Kp luminosities of BCGs
from low-L$_{x}$ clusters are fainter, on average, than the
luminosities of high-L$_{x}$ BCGS. The low-L$_{x}$ BCG
luminosities are most consistent with no-evolution model
predictions. Our confirmation of the observational results of 
previous work suggests, in accordance with previous estimates (ABK), that
these BCGs may be in the process 
of gaining a factor of 2 to 4 in mass via accretion since $z \sim 1$.

\noindent 3)  Tracing the luminosities of high-L$_{x}$ BCGs, we find
they are better described by passive evolution models with a high
redshift of formation, $z_{form}>5$. Again our agreement with previous
studies (BCM) indicates these galaxies are accreting more slowly, 
a factor of 2 in
mass, at most, since $z \sim 1$.

\noindent 4) Combining the low-L$_x$ and high-L$_x$ results, we conclude
that we are seeing evidence for the mass-dependent evolution rates
predicted in hierarchical models. The structures in the more massive
clusters are closer to reaching their final state, and so are 
proportionally accreting more slowly, while the lower mass clusters
are still in a more vigorous state of dynamical evolution.

\noindent 5) From the colors of BCGs ($V-I$ and $I-$\Kp), we
determine that a single evolutionary model is sufficient to describe
the observed color evolution of the stellar populations
of \textit{all} BCGs. The colors of the
combined sample of high-L$_{x}$ and low-L$_{x}$ BCGs are most
consistent with the predictions of passive evolution models with
$z_{form}>5$. 
We find no evidence in the BCG colors for
induced star formation and conclude that the dominant form of
accretion by BCGs since $z \sim 1$ consists of gas-poor, old stellar
systems (stellar populations similar to those already present in the 
BCG).

Because the evolutionary
history of BCGs appears to be linked to properties of the host
cluster, it is significant that we confirm previously observed trends
with a large sample of BCGs that have completely different selection criteria. 
Unfortunately, our photometric data only allow us to speculate on the
merging history of BCGs. Direct evidence of accretion events can be
obtained only with spectroscopy or high resolution
imaging. Spectroscopy will yield
detailed information about the current
star formation in the galaxy and its kinematics but is available for
only a limited number of galaxies at these redshifts (cf. van Dokkum
\& Franx 1996, Kelson \etal 2000, Kelson \etal 2001).
High resolution imaging will complement the spectroscopy
because signatures of merging events include an increase in the effective radius and the slope of the
surface brightness profile of the BCGs (Nelson \etal 2001b, van Dokkum
\& Franx 2001). 
Tracing the structural parameters with
redshift will give an indication of the accretion rate of BCGs. The
combination of a large cluster sample, such as the one presented here,
and spectroscopy or high resolution imaging has the potential to
directly address the question of the merging history of BCGs since
$z \sim 1$.

\vskip 1in
\noindent
Acknowledgments: The authors wish to thank Caryl Gronwall for
providing the GISSEL96 models and offering valuable advice on its
implementation, and Scott Trager for providing valued
comments. AEN and AHG acknowledge funding from a NSF grant
(AST-9733111). AEN gratefully acknowledges financial support from the
University of California Graduate Research Mentorship Fellowship
program. AHG acknowledges funding from the ARCS Foundation and
the CfA postdoctoral fellowship program. DZ acknowledges financial
support from a NSF CAREER grant (AST-9733111) a David and Lucile
Packard Foundation Fellowship, and an Alfred P. Sloan Fellowship. This
work was partially supported by NASA through grant number
GO-07327.01-96A from the Space Telescope Science Institute, which is
operated by the Association of Universities for Research in Astronomy,
Inc., under NASA contract NAS5-26555.

\vfill\eject
\clearpage

\begin{figure*}
\figurenum{1}
\plotone{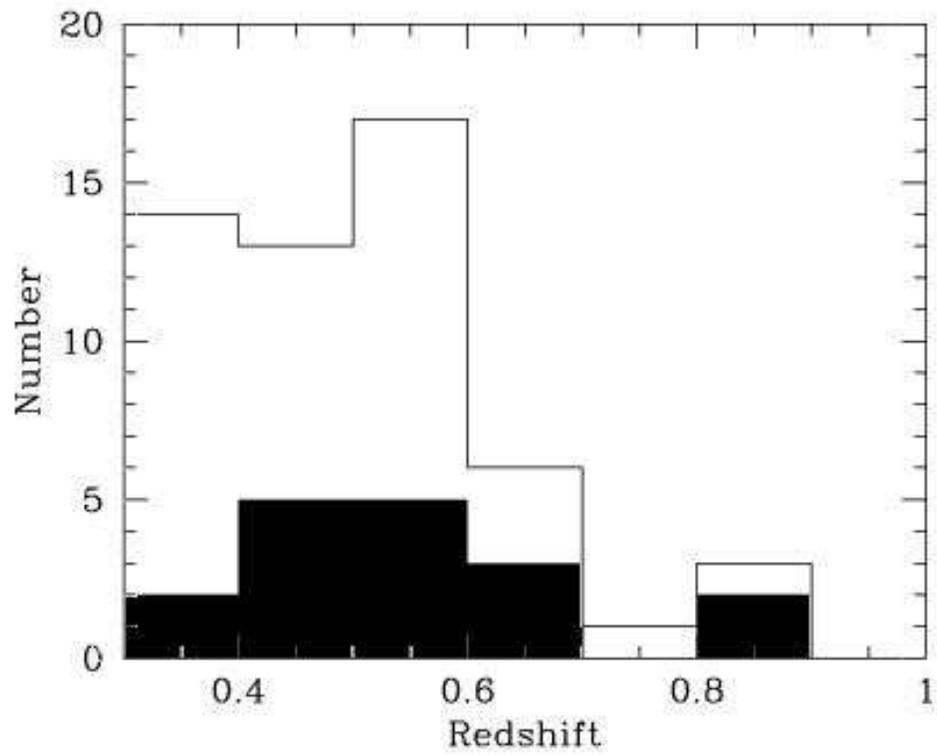}
\protect\figcaption{The redshift distribution of our sample of BCGs with
spectroscopic redshifts (\textit{filled histogram}) and the full
sample (\textit{open histogram}).}
\label{f1} 
\end{figure*}
\clearpage

\begin{figure*}
\figurenum{2}
\plotone{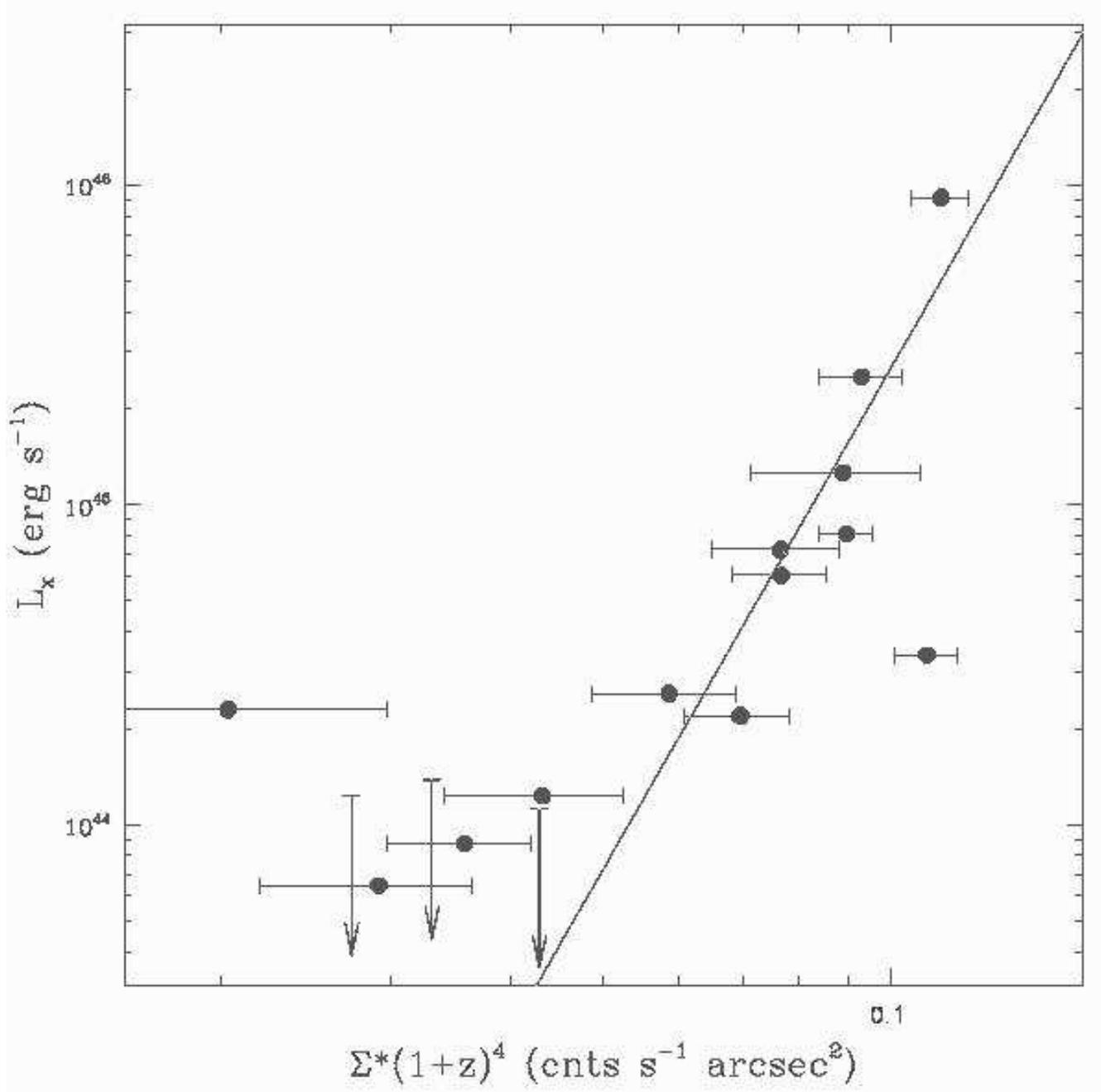}
\protect\figcaption{Comparison of L$_{x}$ with $\Sigma(1 + z)^{5}$ for 17 known x-ray clusters for which we have
drift scan data. L$_{x}$ is the cluster x-ray luminosity in the EMSS
band (0.3-3.5 keV) and $\Sigma$ is the peak surface brightness of
the detection fluctuation. The line is the least squares fit to the
data. 
\label{f2}} 
\end{figure*}
\clearpage

\begin{figure*}
\figurenum{3}
\plotone{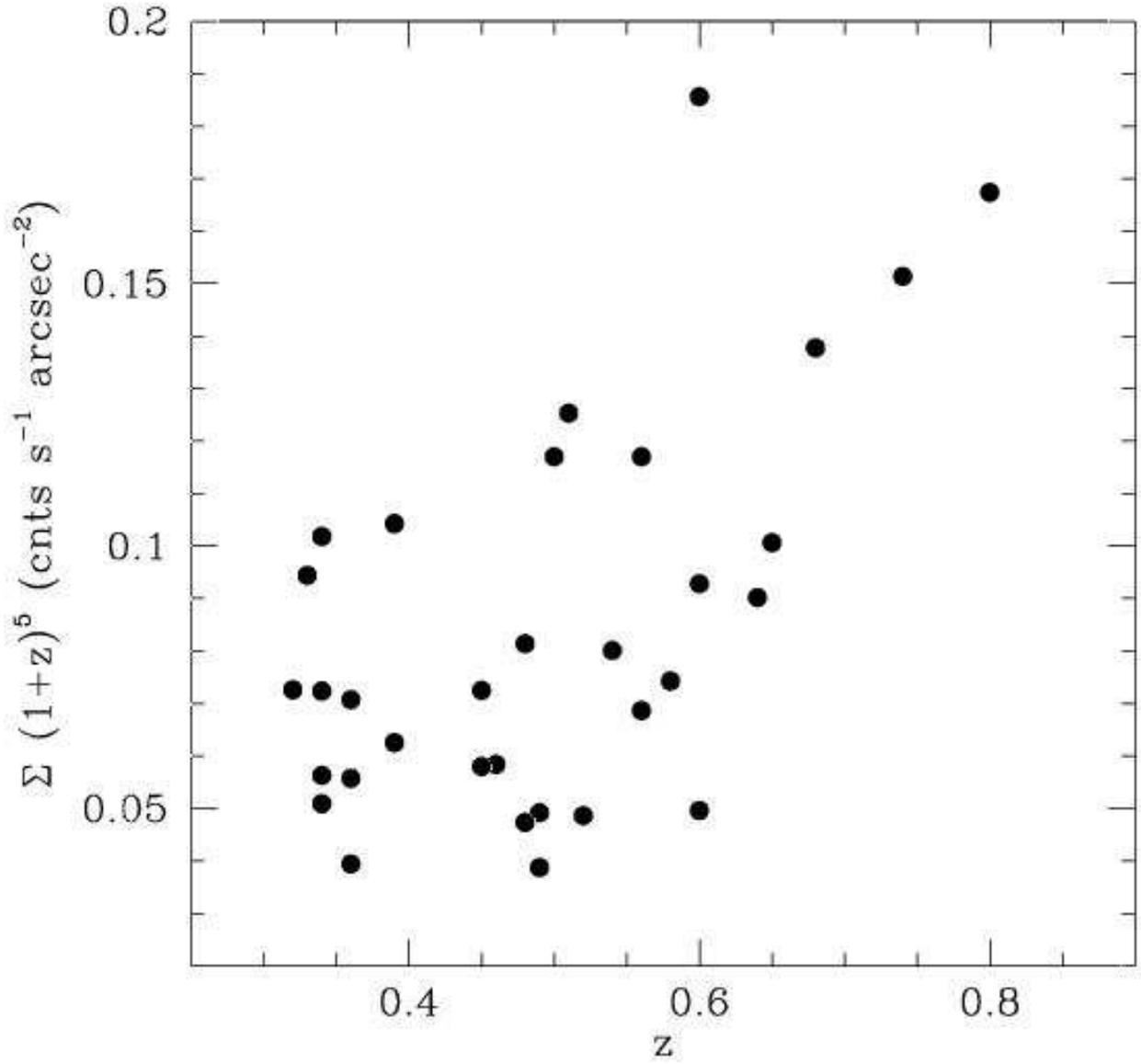}
\protect\figcaption{Comparison of $\Sigma(1 + z)^{5}$ with $z$ for clusters
used in this work. Because $\Sigma(1 + z)^{5}$ correlates with
L$_{x}$, our sample is biased toward more massive systems at high
redshift. 
\label{f3}} 
\end{figure*}
\clearpage

\begin{figure*}
\figurenum{4}
\plotone{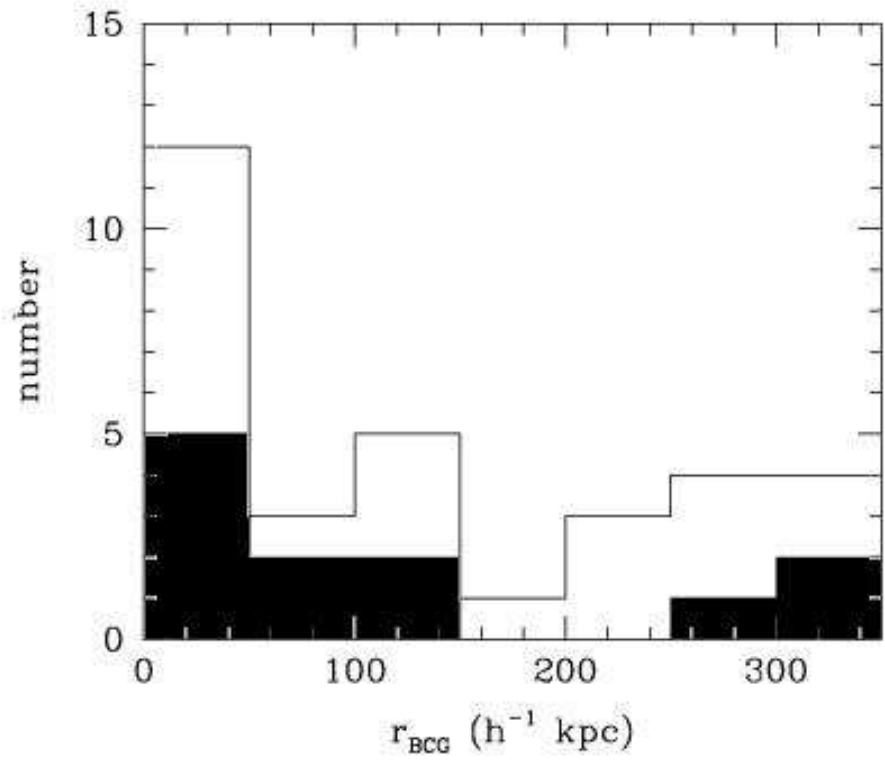}
\protect\figcaption{The distribution of the radial separation of the BCG from
the cluster center, $r_{BCG}$, for all color-selected BCGs
(\textit{open histogram}) and for those with $z_{spec}$ only
(\textit{filled histogram}). 
\label{f4}}
\end{figure*}
\clearpage

\begin{figure*}
\figurenum{5}
\plotone{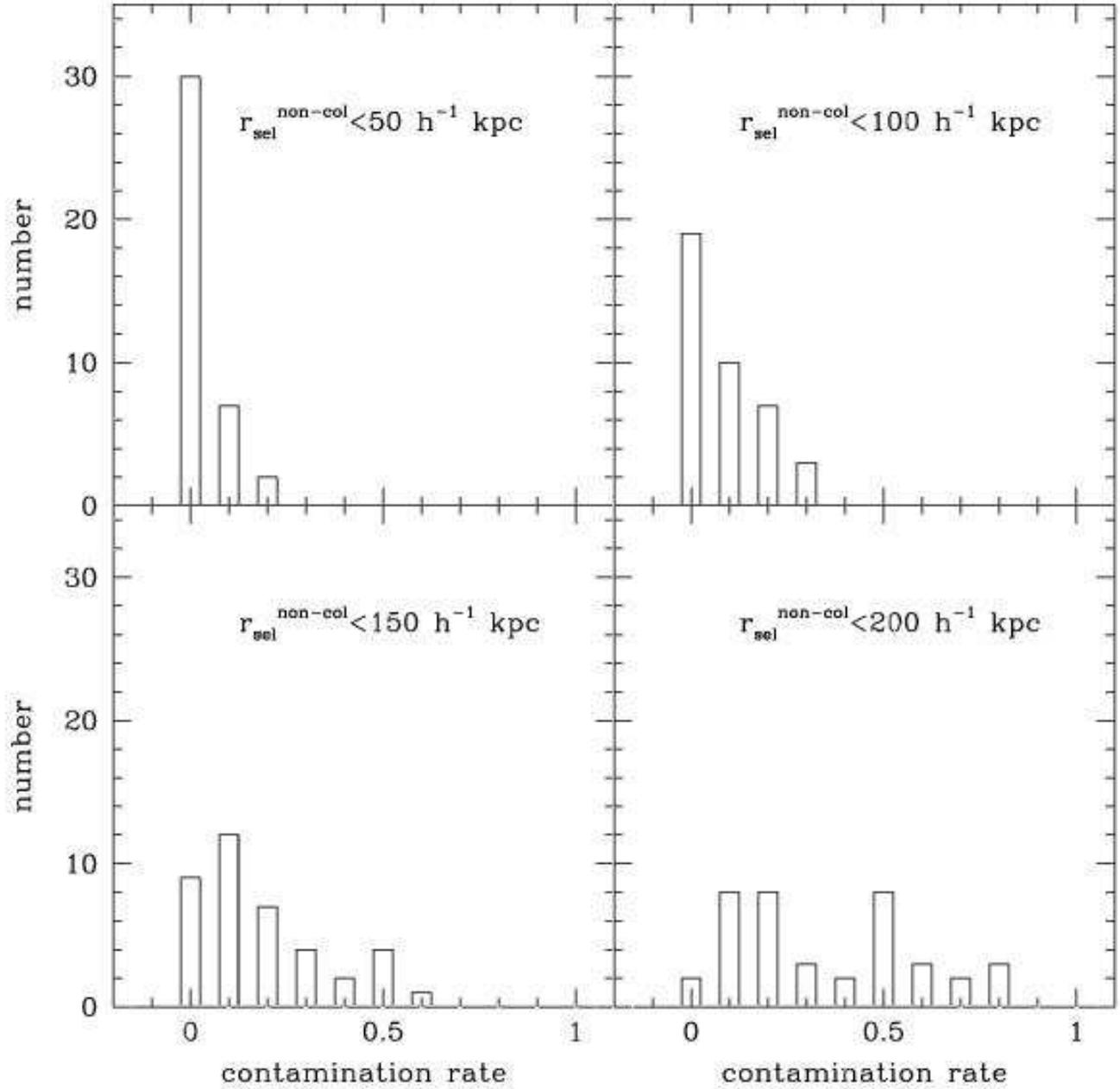}
\protect\figcaption{The distribution of contamination rates for four searching
radii: $r^{non-col}_{sel} = 50 h^{-1}$ kpc (\textit{upper left
panel}), $r^{non-col}_{sel} = 100 h^{-1}$ kpc (\textit{upper right
panel}), $r^{non-col}_{sel} = 150 h^{-1}$ kpc (\textit{lower left
panel}), and $r^{non-col}_{sel} = 200 h^{-1}$ kpc (\textit{lower
right panel}). The contamination rate is defined as the fraction of
times that a galaxy within the search radius is brighter than the
color-selected BCG (\textit{see text for details}).
\label{f5}}
\end{figure*}
\clearpage

\begin{figure*}
\figurenum{6}
\plotone{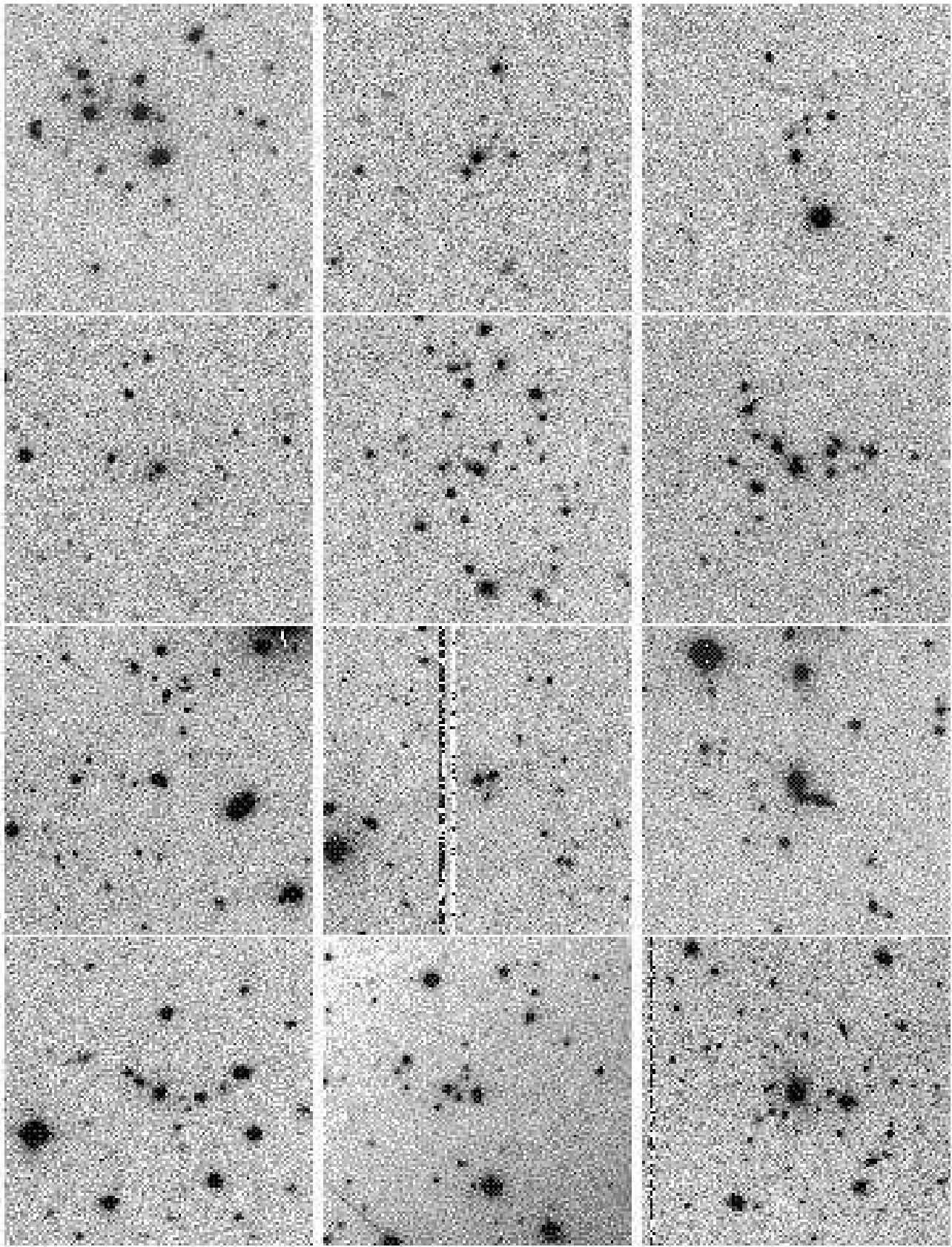}
\protect\figcaption{Postage stamp images of the BCGs used in this work. All
images are in the $I$-band except 1100+4620 which is imaged in
\Kp. The orientation of the images is such that north points up and
east to the right. The images are of (\textit{starting at the top, from left
to right}): 0915+4738, 0936+4620, 1002-1245, 1002-1247, 1005-1209,
1007-1208, 1012-1243, 1012-1245, 1014-1143, 1015-1132, 1017-1128,
1018-1211. (Due to size restrictions 
the astro-ph version only shows one of the set of five panels. The others
are available at ngala.as.arizona.edu:/dennis/clusters.html or the
published version)
\label{f6a}}
\end{figure*}
\clearpage

\begin{figure*}
\figurenum{7}
\plotone{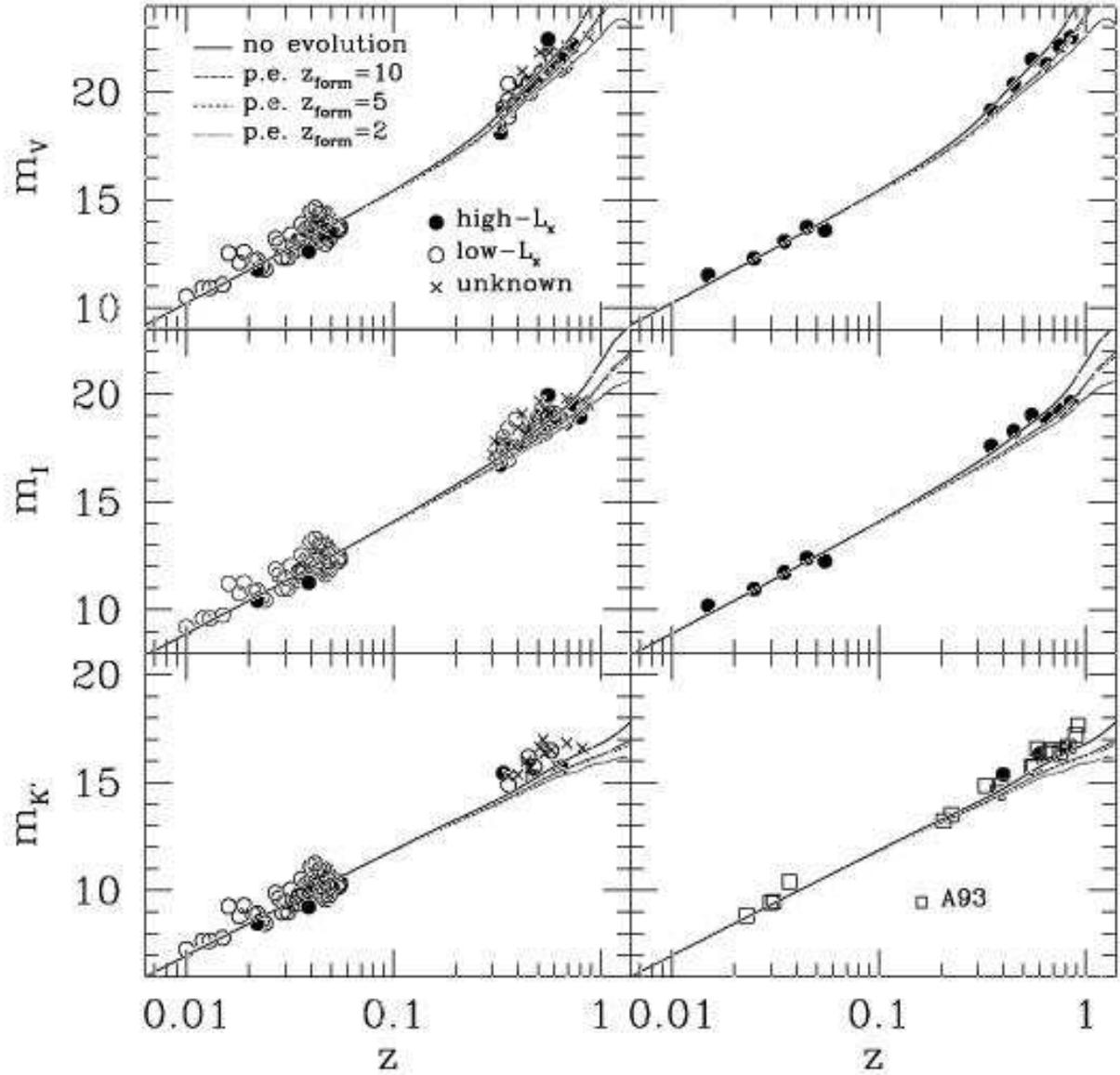}
\protect\figcaption{The $V$-band (\textit{upper panels}), $I$-band
(\textit{center panels}), and $K^{\prime}$-band (\textit{lower
panels}) Hubble diagrams for BCGs from high-L$_{x}$ clusters
(\textit{left panels, filled circles}), low-L$_{x}$ clusters
(\textit{left panels, open circles}), and clusters for which L$_{x}$
cannot be estimated (\textit{left panels, crosses}). The threshold
between high-L$_{x}$ and low-L$_{x}$ is $2 \times 10^{44}$ ergs
s$^{-1}$. The errors in the photometry are negligible, but an 
error in $z_{phot}$ of 0.07 corresponds to $\Delta m_I \sim 0.3$ mag
at $z \sim 0.6$.
The BCG luminosities are compared to the spectral synthesis
model predictions of Bruzual \& Charlot (1993) for an elliptical
galaxy experiencing passive evolution and no-evolution. The low
redshift points are BCGs from PL95 for which the host cluster has
x-ray data and are used to normalize the models. In the right panels,
our data are binned in $\Delta z = $0.1 for the $V$- and $I$-bands
(while $\Delta z = 0.01$ for PL95). For the \Kp-band we bin the data
in $\Delta z = 0.2$ and compare to the $K$-band BCG sample of A93
(\textit{open squares}), which is not binned. The error bars, which
are $\sigma_{mean}$ of the binned data, are smaller than the points
and are omitted.  \label{f7}}
\end{figure*}
\clearpage

\begin{figure*}
\figurenum{8}
\plotone{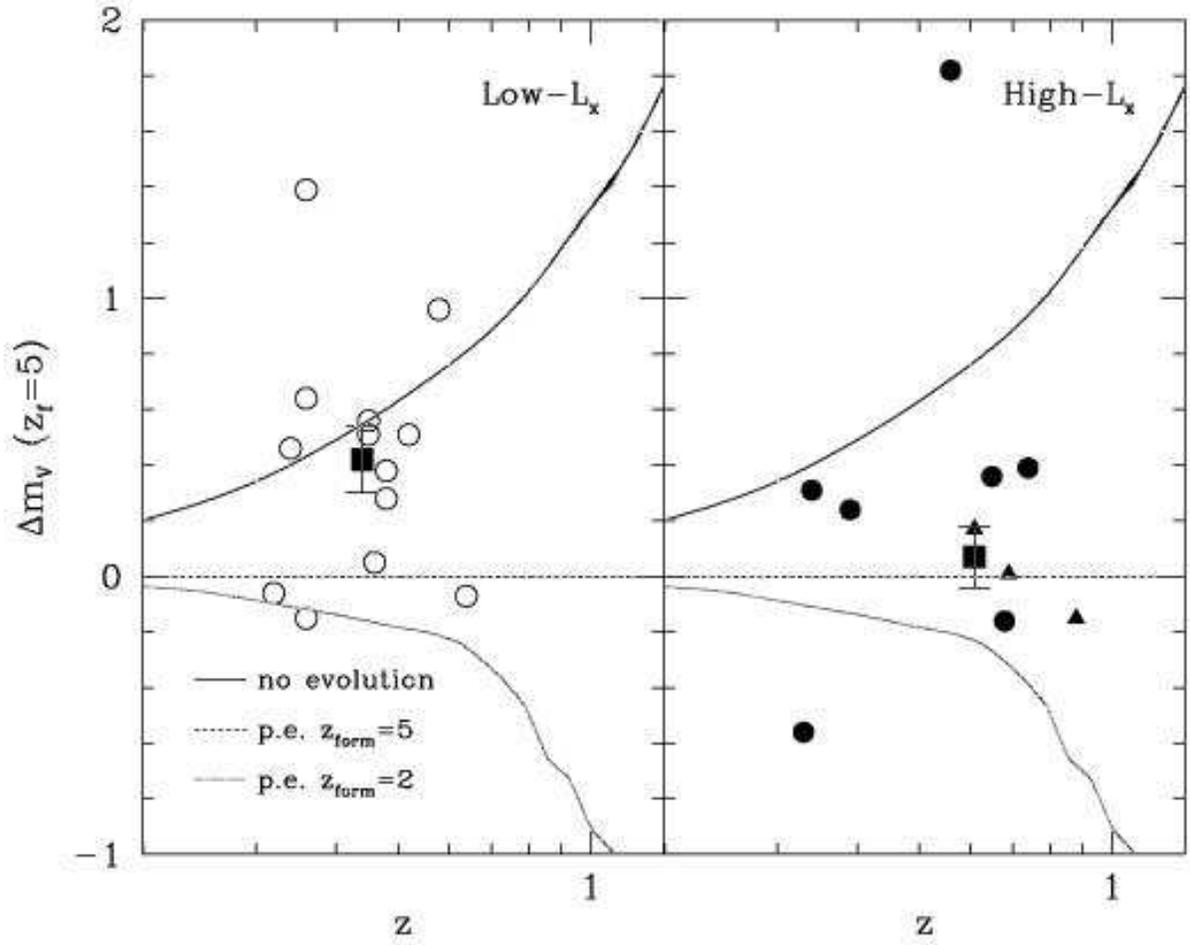}
\protect\figcaption{BCG $V$ magnitude residuals for low-L$_{x}$ clusters
(\textit{left panel}) and high-L$_{x}$ clusters (\textit{right
panel}). Because we detect only very massive clusters at high
redshift, clusters without measured values of $\Sigma$ at $z \geq 0.6$
are included in the high-L$_{x}$ sample (\textit{filled triangles}). Residuals are defined as $\Delta m = m_{BCG} - m_{model}$,
for a passive evolution model with $z_{form} = 5$. 
The filled boxes are the mean magnitude residuals and have error bars that are $\sigma_{mean}$.
The lines are the same as those in Figure 7.
\label{f8}} 
\end{figure*}
\clearpage

\begin{figure*}
\figurenum{9}
\plotone{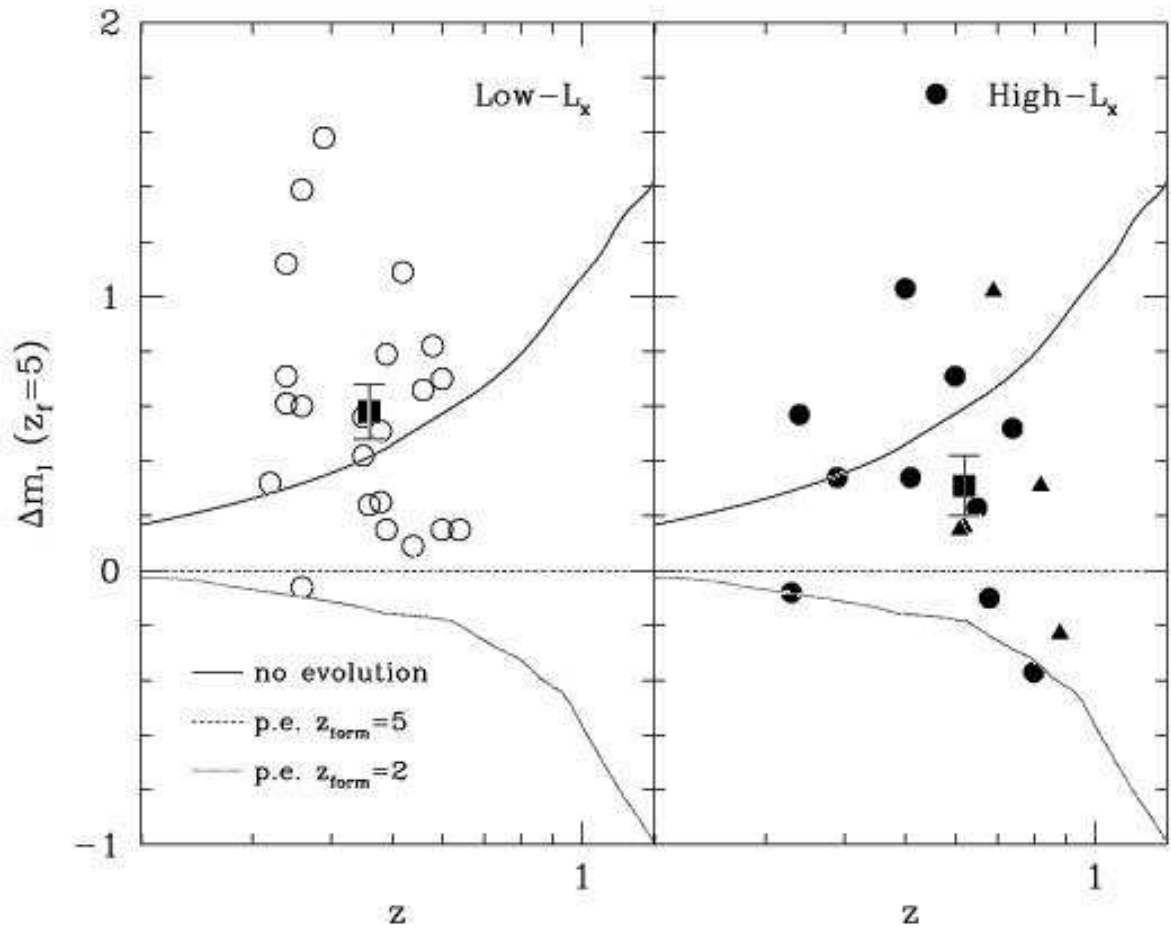}
\protect\figcaption{Same as for Figure 8 but in the $I$-band}
\label{f9}
\end{figure*}
\clearpage

\begin{figure*}
\figurenum{10}
\plotone{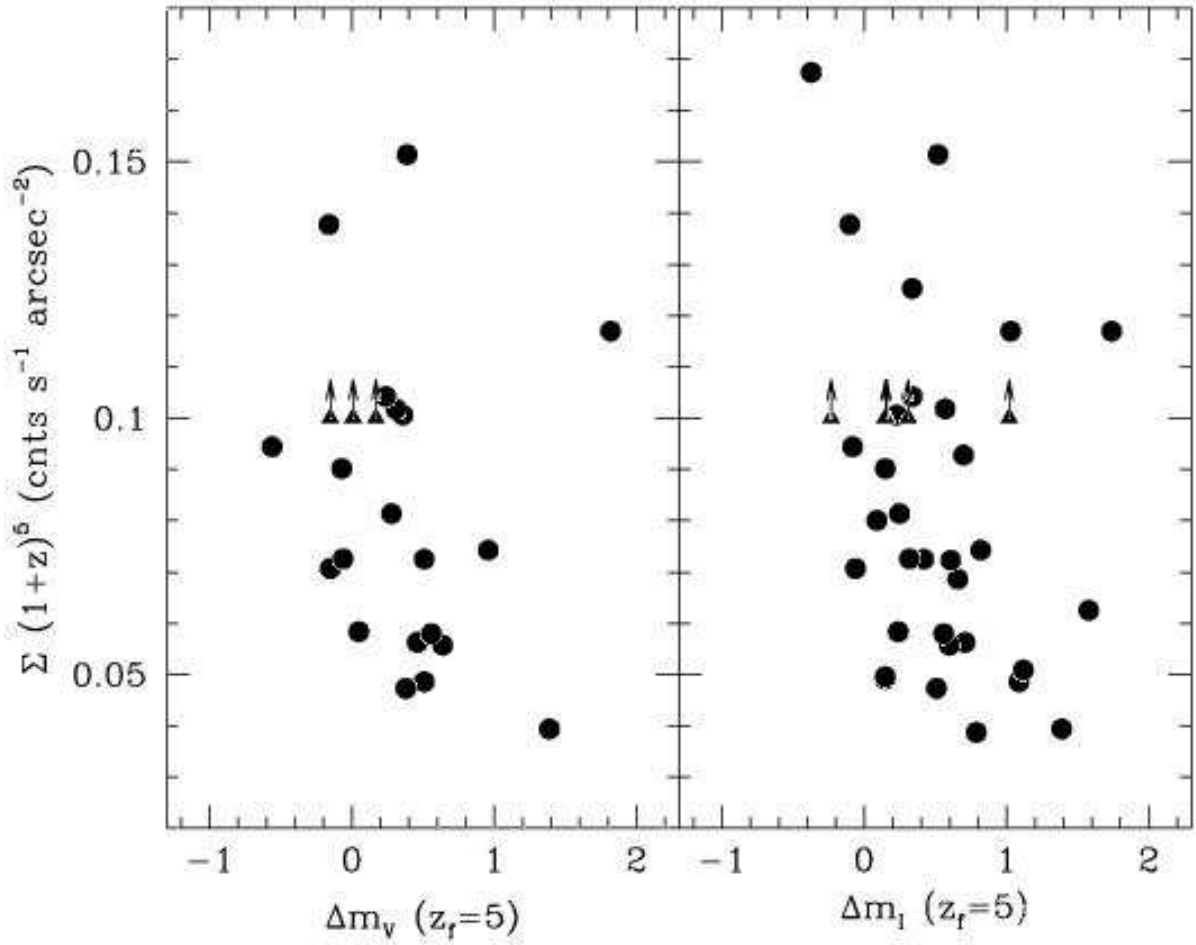}
\protect\figcaption{Comparison of BCG magnitude residuals, $\Delta m = m_{BCG} - m_{model}$, in $V$
(\textit{left panel}) and $I$ (\textit{right panel}) with
$\Sigma(1 + z)^{5}$ for passive evolution with $z_{form} = 5$. Filled
triangles are BCGs at $z \geq 0.6$ without measured values of
$\Sigma$. Because we are only sensitive to massive clusters at high
redshift, we assign these BCGs lower limiting values of $\Sigma(1 + z)^{5} = 0.100$.
\label{f10}}
\end{figure*}
\clearpage

\begin{figure*}
\figurenum{11}
\plotone{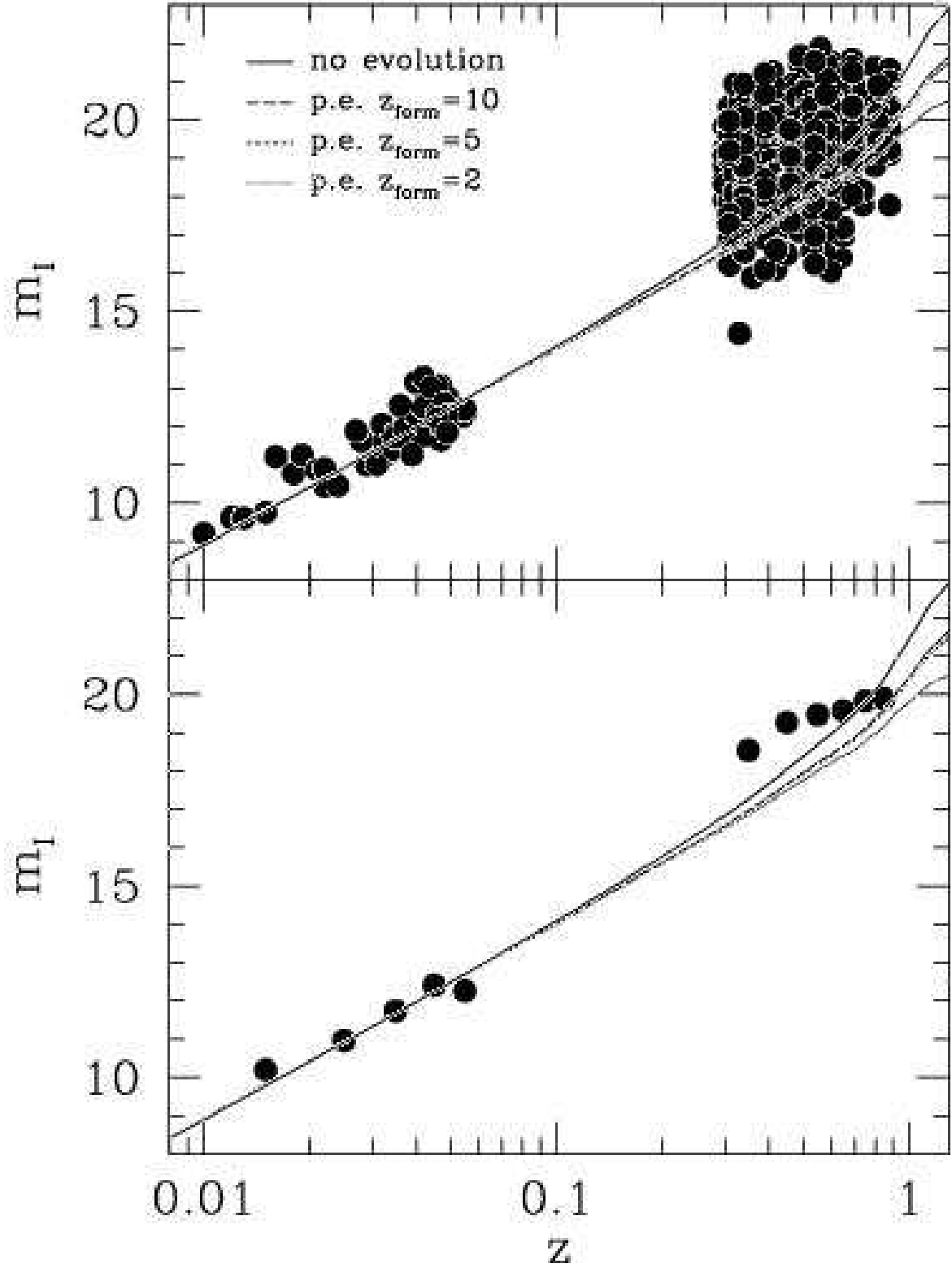}
\protect\figcaption{$I$ Hubble diagram for the brightest field galaxies
selected in random areas comparable to that of our clusters
(\textit{upper panel}). In the lower panel, the data are
binned in $\Delta z = 0.1$. The error bars, which are $\sigma_{mean}$ of the binned
data, are smaller than the points and are omitted.
\label{f11}}
\end{figure*}
\clearpage

\begin{figure*}
\figurenum{12}
\plotone{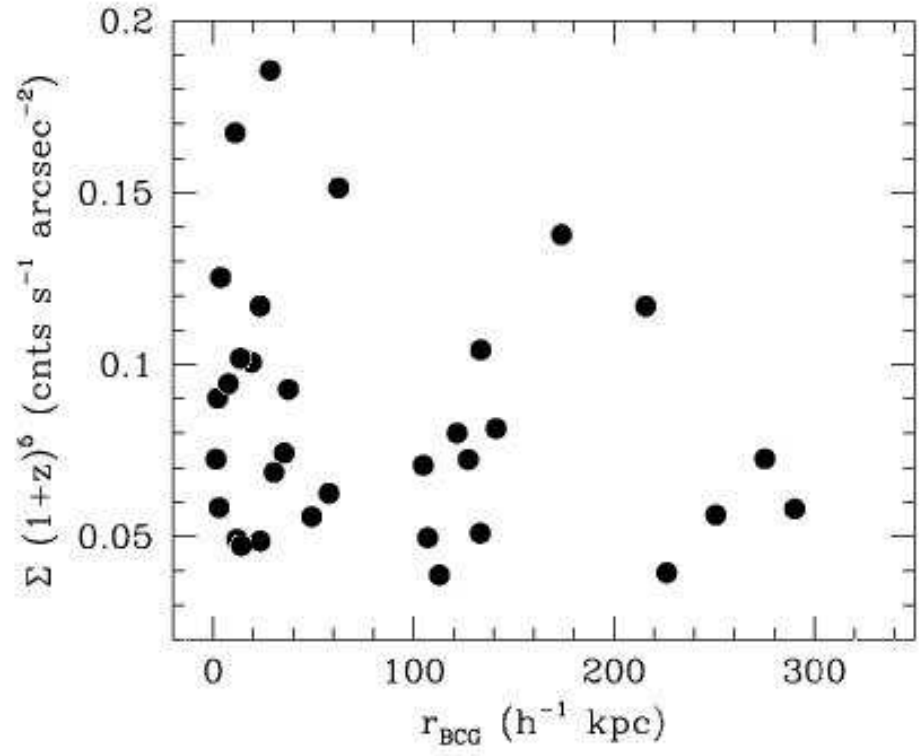}
\figcaption{Comparison of $\Sigma(1 + z)^{5}$ with $r_{BCG}$. The lack
of a statistically significant correlation
between $\Sigma(1 + z)^{5}$ and $r_{BCG}$ suggests that
the measurements of $\Sigma$ are not biased by contamination from the
BCG halo. 
\label{f12}}
\end{figure*}
\clearpage

\begin{figure*}
\figurenum{13}
\plotone{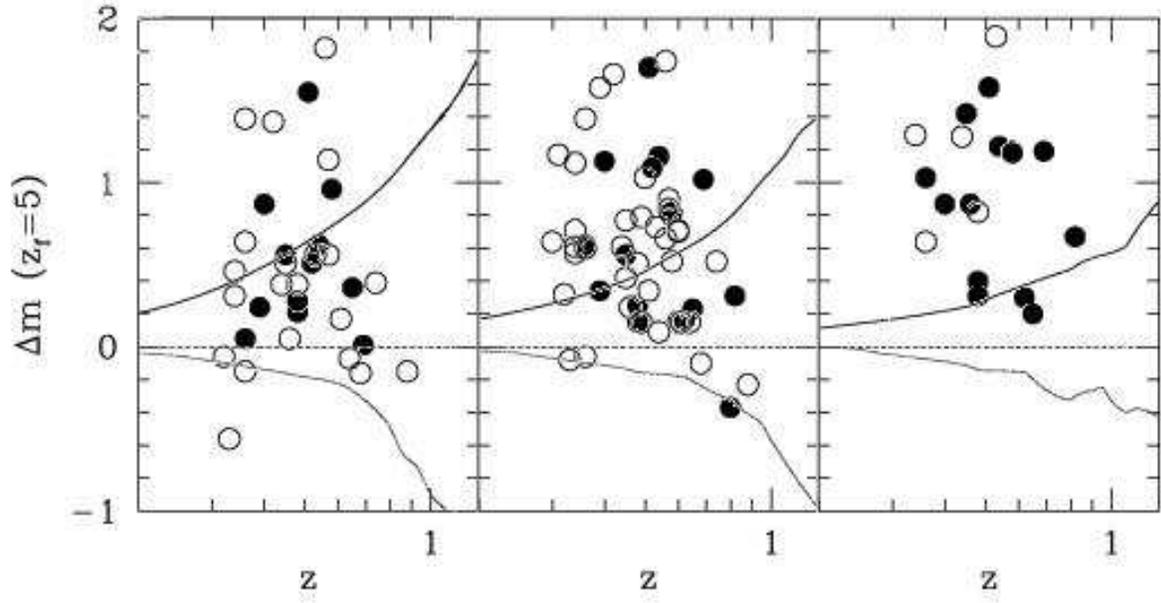}
\protect\figcaption{BCG magnitude residuals in $V$ (\textit{left panel}), $I$
(\textit{center panel}), and \Kp (\textit{right panel}) for clusters
with $z_{spec}$ (\textit{filled circles}) and $z_{phot}$ (\textit{open
circles}). Residuals are defined as $\Delta m = m_{BCG} - m_{model}$,
for a passive evolution model with $z_{form} = 5$. The lines are the same as
those in Figure 7.
\label{f13}}
\end{figure*}
\clearpage

\begin{figure*}
\figurenum{14}
\plotone{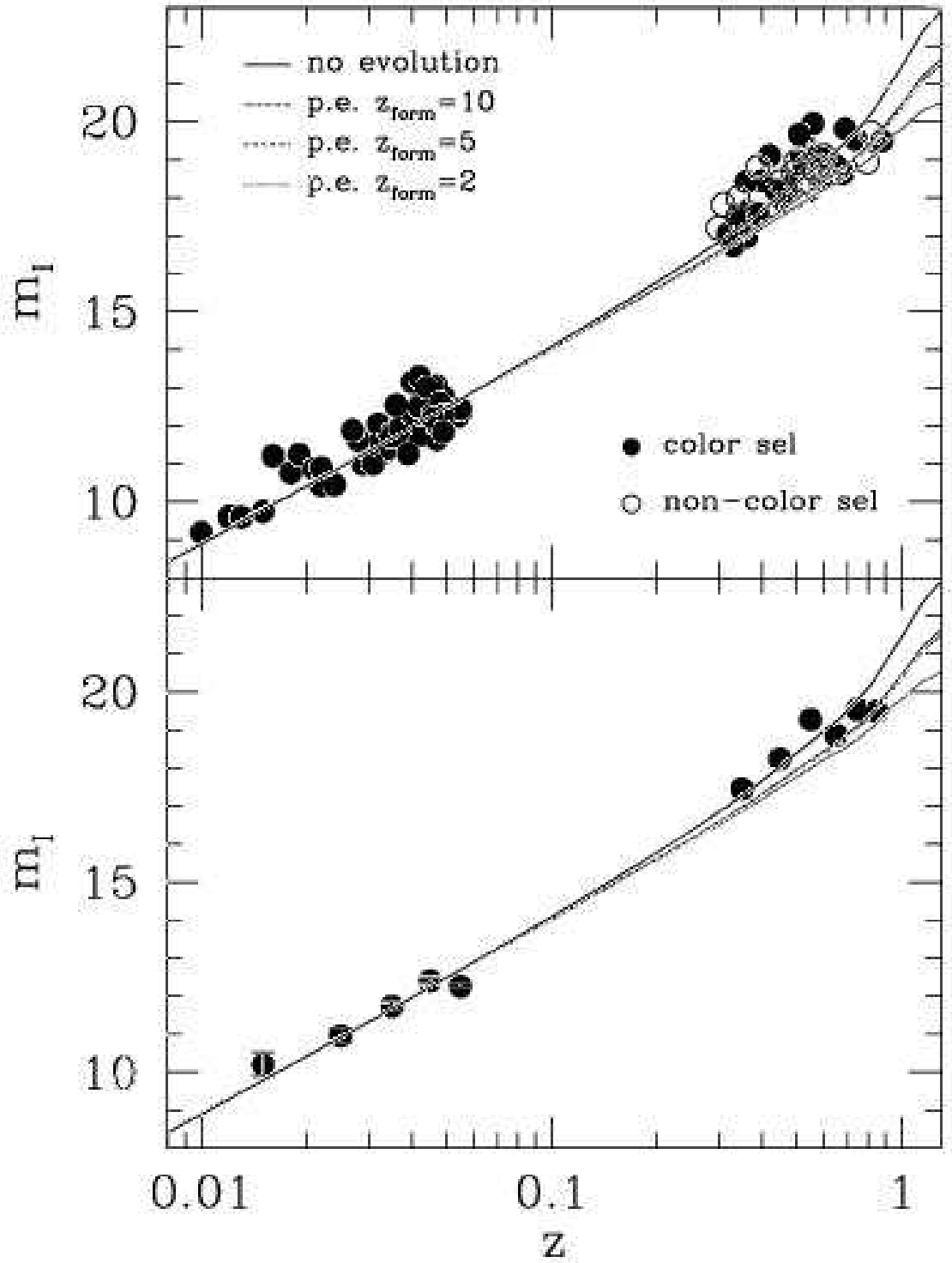}
\protect\figcaption{The $I$-band Hubble diagram from Figure 7,
differentiating between color-selected and non-color-selected
BCGs. Symbols are the same as for Figure 7.
\label{f14}}
\end{figure*}
\clearpage

\begin{figure*}
\figurenum{15}
\plotone{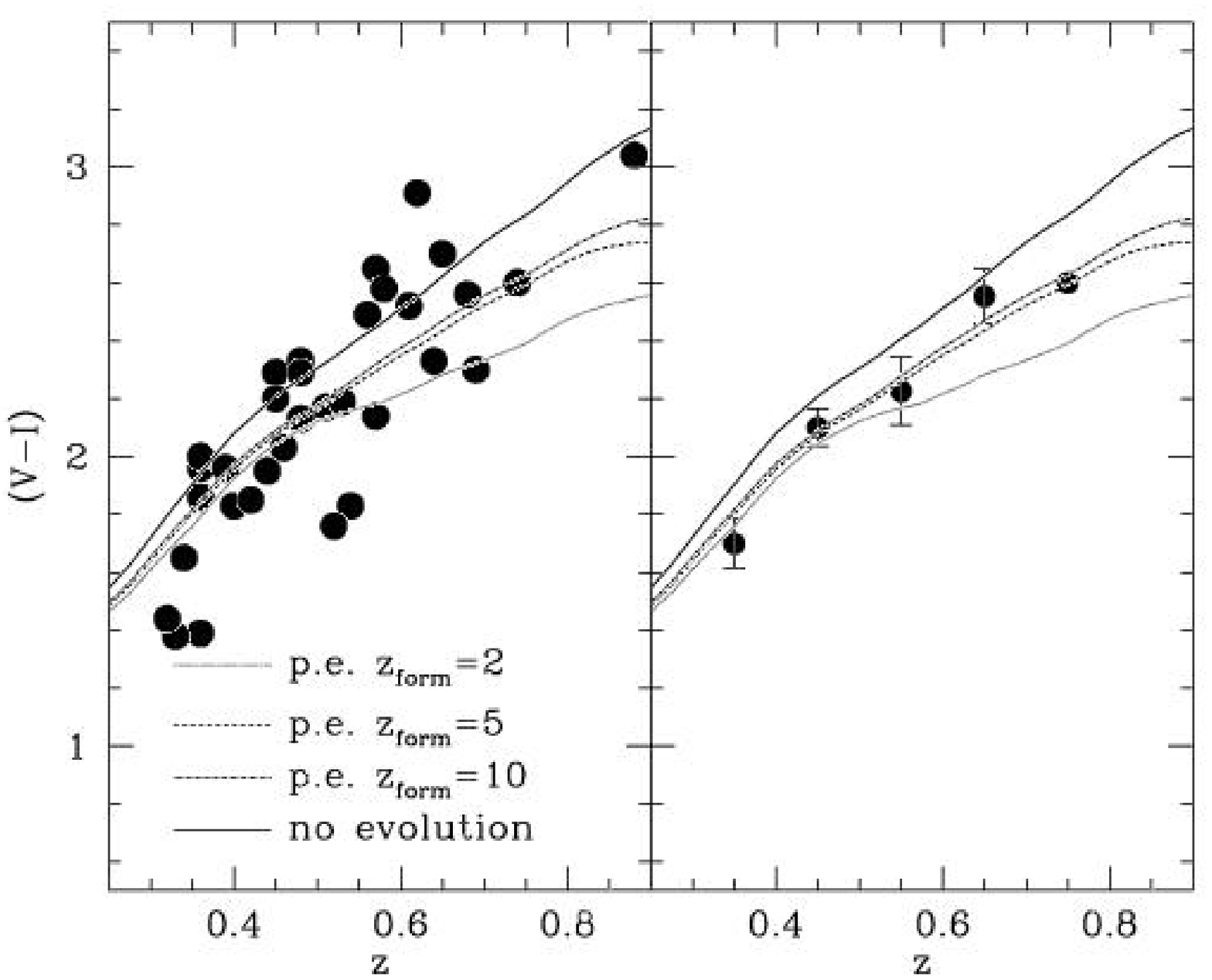}
\protect\figcaption{BCG colors in $V-I$ compared to the spectral synthesis
model predictions of Bruzual \& Charlot (1993) for an elliptical galaxy
experiencing passive evolution and no-evolution (\textit{left
panel}). In the right panel, the data are binned
in $\Delta z = 0.1$ and have error bars which are $\sigma_{mean}$ of
the binned data.
\label{f15}} 
\end{figure*}
\clearpage

\begin{figure*}
\figurenum{16}
\plotone{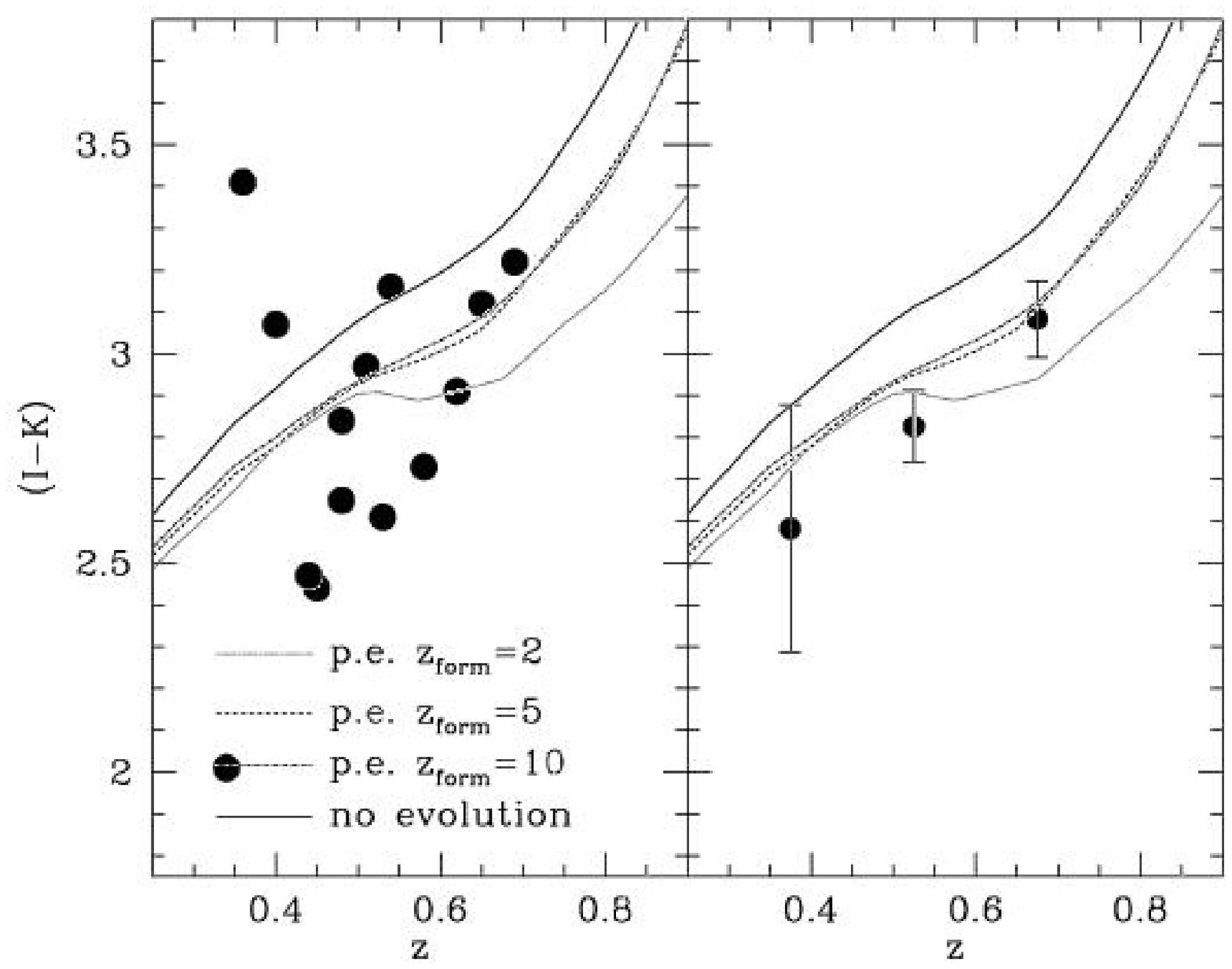}
\protect\figcaption{BCG colors in $I-$\Kp compared to the spectral synthesis
model predictions of Bruzual \& Charlot (1993) for an elliptical galaxy
experiencing passive evolution and no-evolution (\textit{left
panel}). In the right panel, the data are binned
in $\Delta z = 0.15$ and have error bars which are $\sigma_{mean}$ of
the binned data.
\label{f16}}
\end{figure*}
\clearpage
 
\begin{figure*}
\figurenum{17}
\plotone{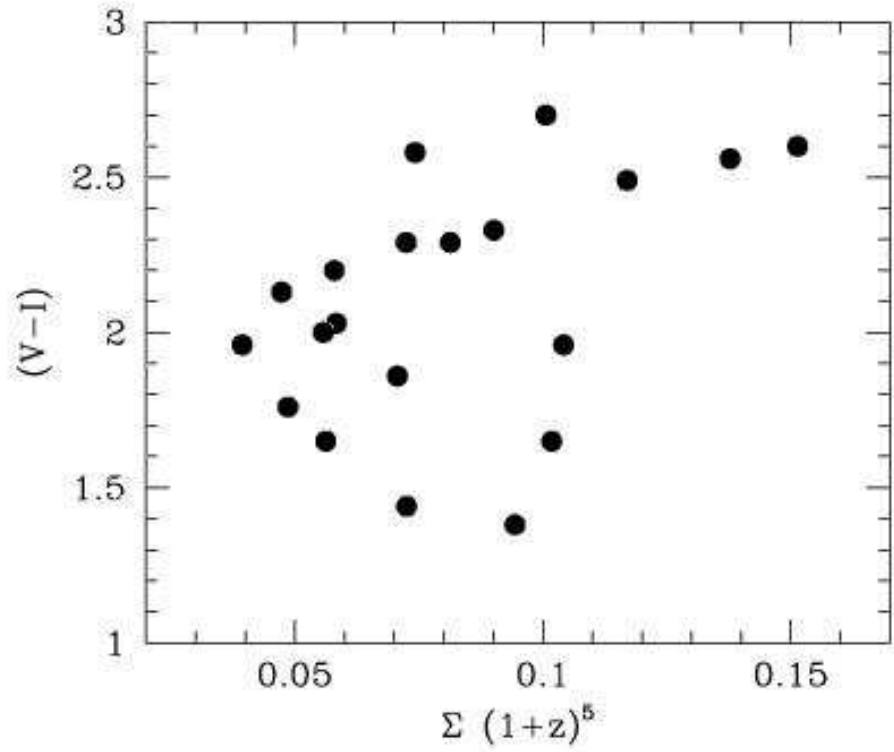}
\protect\figcaption{Comparison of the colors of BCGs in $V-I$ with
$\Sigma(1 + z)^{5}$. There is no statistically significant correlation
between the BCG colors and $\Sigma(1 + z)^{5}$.
\label{f17}}
\end{figure*}
\clearpage

\begin{figure*}
\figurenum{18}
\plotone{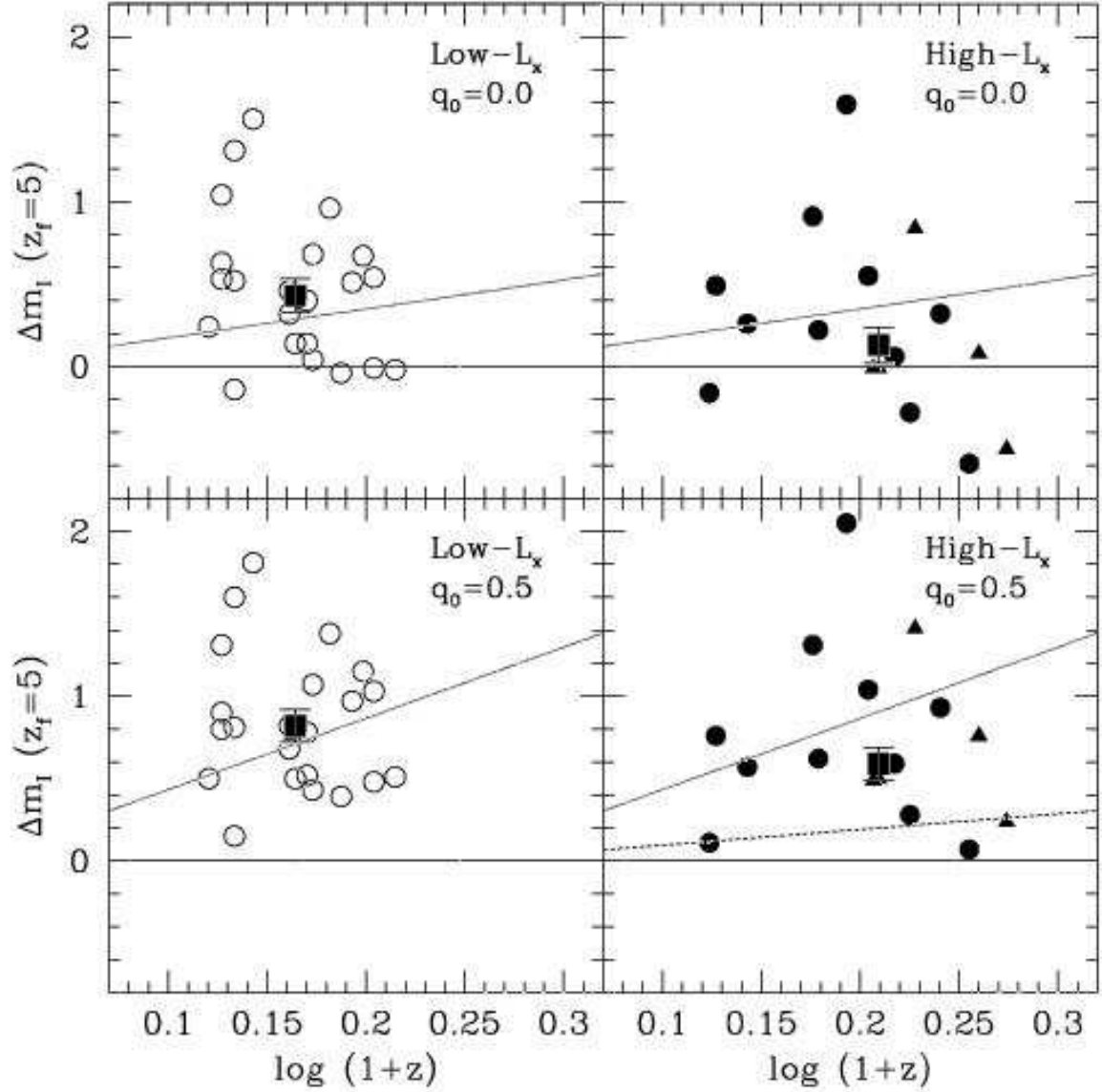}
\protect\figcaption{BCG $I$ magnitude residuals for low-L$_{x}$ clusters
(\textit{left panels}) and high-L$_{x}$ clusters (\textit{right
panels}). Residuals are defined as $\Delta m = m_{BCG} - m_{model}$,
for a passive evolution model with $z_{form} = 5$ (\textit{solid
line}) assuming $\Omega_{m} = 0.0, \Omega_\Lambda=0.0$ (\textit{upper
panels}) and $\Omega_{m} = 1, \Omega_\Lambda=0$ (\textit{lower
panels}) and H$_{0} = $50 km s$^{-1}$ Mpc$^{-1}$. Symbols are the same
as for Figure 8. The filled boxes denote the mean magnitude residuals
and have errorbars which are $\sigma_{mean}$ of the data. The dotted
lines are the mass accretion rates predicted by ABK for low-L$_{x}$
BCGs, while the dashed line is the mass accretion rate from BCM for
high-L$_{x}$ BCGs.
\label{f18}}
\end{figure*}
\clearpage

%\documentclass[preprint]{aastex}
%\pagestyle{empty}
%def\etal{et al.}
%\begin{document}
\begin{deluxetable}{cccccccccccc}
%\tabletypesize{\small}%\tablewidth{0pc}
\tablenum{1}
\tablecaption{Observations}
\tablewidth{435pt}
\tablehead{
\colhead{} & \colhead{RA} & \colhead{DEC} & \colhead{V} & \colhead{Seeing} & \colhead{I} & \colhead{Seeing} & \colhead{K} & \colhead{Seeing} \\ 
\colhead{Cluster} & \colhead{(JD2000)} & \colhead{(JD2000)} & \colhead{(min)} & \colhead{($^{\prime\prime}$)} & \colhead{(min)} & \colhead{($^{\prime\prime}$)} & \colhead{(min)} & \colhead{($^{\prime\prime}$)} }

 \startdata 
0915+4738 & 09:15:51.94 & +47:38:19.9 & 50\tablenotemark{2} & 1.8 & 33\tablenotemark{2} & 1.9 & 56\tablenotemark{2} & 2.0\\
0936+4620 & 09:36:06.28 & +46:20:43.2 & 80\tablenotemark{2} & 2.5 & 80\tablenotemark{2} & 1.9 & 64\tablenotemark{2} & 2.0\\
0944+4732 & 09:44:21.83 & +47:32:42.6 & ... & ... & ... & ... & 40\tablenotemark{4} & 2.0\\
1002$-$1245 & 10:02:01.47 & $-$12:45:35.1 & 120\tablenotemark{1} & 1.4 & 60\tablenotemark{1} & 1.4 & ... & ...\\
1002$-$1247 & 10:02:27.14 & $-$12:47:13.1 & 120\tablenotemark{1} & 1.4 & 60\tablenotemark{1} & 1.4 & ... & ...\\
1005$-$1147 & 10:05:43.60 & $-$11:47:43.1 & ... & ... & 60\tablenotemark{1} & 1.2 & ... & ...\\
1005$-$1209 & 10:05:49.72 & $-$12:09:36.5 & ... & ... & 45\tablenotemark{1} & 1.4 & ... & ...\\
1006$-$1222 & 10:06:29.25 & $-$12:22:13.7 & ... & ... & 45\tablenotemark{1} & 1.4 & ... & ...\\
1006$-$1258 & 10:06:18.79 & $-$12:58:12.5 & ... & ... & 68\tablenotemark{1} & 1.1 & ... & ...\\
1007$-$1208 & 10:07:42.60 & $-$12:08:36.0 & ... & ... & 40\tablenotemark{1} & 1.3 & ... & ...\\
1012$-$1243 & 10:12:14.58 & $-$12:43:10.9 & ... & ... & 45\tablenotemark{3} & 0.8 & 32\tablenotemark{3} & 0.8\\
1012$-$1245 & 10:12:44.35 & $-$12:45:37.9 & 30\tablenotemark{3} & 0.9 & 30\tablenotemark{3} & 0.7 & ... & ...\\
1014$-$1143 & 10:14:56.31 & $-$11:43:08.1 & ... & ... & 45\tablenotemark{1} & 1.2 & ... & ...\\
1015$-$1132 & 10:15:19.47 & $-$11:32:55.5 & ... & ... & 60\tablenotemark{1} & 1.3 & ... & ...\\
1017$-$1128 & 10:17:45.31 & $-$11:28:07.8 & ... & ... & 60\tablenotemark{3} & 1.1 & ... & ...\\
1018$-$1211 & 10:18:46.45 & $-$12:11:52.8 & 40\tablenotemark{3} & 1.2 & 40\tablenotemark{3} & 0.6 & ... & ...\\
1023$-$1303 & 10:23:10.11 & $-$13:03:52.0 & ... & ... & 30\tablenotemark{3} & 0.6 & ... & ...\\
1024$-$1239 & 10:24:44.94 & $-$12:39:55.6 & 40\tablenotemark{3} & 1.1 & 40\tablenotemark{3} & 0.9 & ... & ...\\
1025$-$1236 & 10:25:08.83 & $-$12:36:20.0 & 90\tablenotemark{1} & 1.5 & 115\tablenotemark{1} & 2.7 & ... & ...\\
1027$-$1159 & 10:27:26.31 & $-$11:59:33.6 & 90\tablenotemark{1} & 1.5 & 105\tablenotemark{1} & 1.8 & ... & ...\\
1031$-$1244 & 10:31:50.26 & $-$12:44:27.2 & 40\tablenotemark{3} & 1.0 & 60\tablenotemark{3} & 0.8 & ... & ...\\
1032$-$1229 & 10:32:04.91 & $-$12:29:43.8 & 120\tablenotemark{1} & 1.8 & 115\tablenotemark{1} & 1.8 & ... & ...\\
... & ... & ... & ... & ... & 15\tablenotemark{3} & 1.0 & ... & ...\\
1041+4626 & 10:41:03.79 & +46:26:36.3 & 40\tablenotemark{2} & 2.5 & 60\tablenotemark{2} & 2.2 & 118\tablenotemark{2} & 2.0\\
1059+4737 & 10:59:38.03 & +47:37:38.6 & 50\tablenotemark{2} & 3.3 & 50\tablenotemark{2} & 2.0 & 55\tablenotemark{2} & 2.0\\
1100+4620 & 11:00:57.36 & +46:20:38.3 & ... & ... & ... & ... & 88\tablenotemark{2} & 2.0\\
1136$-$1136 & 11:36:31.95 & $-$11:36:07.7 & 30\tablenotemark{3} & 1.3 & 30\tablenotemark{3} & 1.1 & ... & ...\\
... & ... & ... & ... & ... & 20\tablenotemark{1} & 1.1 & ... & ...\\
1136$-$1145 & 11:36:47.07 & $-$11:45:33.3 & 13\tablenotemark{3} & 1.1 & 20\tablenotemark{3} & 0.9 & 44\tablenotemark{3} & 0.8\\
... & ... & ... & 40\tablenotemark{3} & 1.0 & 40\tablenotemark{3} & 0.8 & ... & ...\\
1136$-$1252 & 11:36:33.54 & $-$12:52:03.3 & 20\tablenotemark{3} & 0.8 & 20\tablenotemark{3} & 0.6 & 20\tablenotemark{4} & 2.0\\
... & ... & ... & ... & ... & 20\tablenotemark{3} & 1.4 & ... & ...\\
1138$-$1142 & 11:38:06.59 & $-$11:42:10.3 & 20\tablenotemark{3} & 1.1 & 30\tablenotemark{3} & 0.7 & 32\tablenotemark{3} & 0.8\\
... & ... & ... & 60\tablenotemark{3} & 1.3 & 60\tablenotemark{3} & 1.2 & ... & ...\\
1138$-$1225 & 11:38:14.24 & $-$12:25:53.9 & 90\tablenotemark{1} & 1.2 & 120\tablenotemark{1} & 2.0 & ... & ...\\
1138$-$1228 & 11:38:44.70 & $-$12:28:25.4 & 30\tablenotemark{3} & 0.7 & 20\tablenotemark{3} & 0.8 & ... & ...\\
1139$-$1154 & 11:39:30.47 & $-$11:54:25.0 & 30\tablenotemark{3} & 0.9 & 30\tablenotemark{3} & 0.8 & 10\tablenotemark{4} & 2.0\\
1139$-$1217 & 11:39:56.84 & $-$12:17:19.8 & 20\tablenotemark{3} & 0.7 & 20\tablenotemark{3} & 0.8 & 32\tablenotemark{3} & 0.8\\
1145$-$1155 & 11:45:22.36 & $-$11:55:52.1 & 20\tablenotemark{3} & 0.7 & 20\tablenotemark{3} & 1.0 & 50\tablenotemark{4} & 2.0\\
... & ... & ... & 40\tablenotemark{3} & 1.2 & 30\tablenotemark{3} & 0.8 & ... & ...\\
1147$-$1252 & 11:47:16.98 & $-$12:52:04.7 & 20\tablenotemark{3} & 1.0 & 20\tablenotemark{3} & 0.9 & 32\tablenotemark{3} & 0.8\\
... & ... & ... & 40\tablenotemark{3} & 1.2 & 30\tablenotemark{3} & 0.7 & ... & ...\\
1149$-$1159 & 11:49:22.14 & $-$11:59:19.3 & 3\tablenotemark{3} & 0.9 & 10\tablenotemark{3} & 0.9 & ... & ...\\
1149$-$1246 & 11:49:06.18 & $-$12:46:06.3 & 20\tablenotemark{3} & 0.9 & 20\tablenotemark{3} & 0.8 & 32\tablenotemark{3} & 0.8\\
... & ... & ... & 40\tablenotemark{3} & 1.0 & 40\tablenotemark{3} & 0.7 & ... & ...\\
1208$-$1151 & 12:08:26.67 & $-$11:51:25.6 & 90\tablenotemark{1} & 1.4 & 68\tablenotemark{1} & 1.2 & ... & ...\\
1210$-$1219 & 12:10:12.73 & $-$12:19:06.9 & 60\tablenotemark{1} & 1.4 & 60\tablenotemark{1} & 1.4 & ... & ...\\
1211$-$1220 & 12:11:04.16 & $-$12:20:47.7 & ... & ... & 45\tablenotemark{3} & 0.8 & ... & ...\\
1215$-$1252 & 12:15:41.08 & $-$12:52:59.7 & ... & ... & 68\tablenotemark{1} & 1.1 & ... & ...\\
1216$-$1201 & 12:16:45.10 & $-$12:01:17.3 & ... & ... & 45\tablenotemark{1} & 1.3 & ... & ...\\
\enddata
\end{deluxetable}

\clearpage

\begin{deluxetable}{cccccccccccc}
%\tabletypesize{\small}%\tablewidth{0pc}
\tablenum{1}
\tablecaption{Observations}
\tablewidth{435pt}
\tablehead{
\colhead{} & \colhead{RA} & \colhead{DEC} & \colhead{V} & \colhead{Seeing} & \colhead{I} & \colhead{Seeing} & \colhead{K} & \colhead{Seeing} \\ 
\colhead{Cluster} & \colhead{(JD2000)} & \colhead{(JD2000)} & \colhead{(min)} & \colhead{($^{\prime\prime}$)} & \colhead{(min)} & \colhead{($^{\prime\prime}$)} & \colhead{(min)} & \colhead{($^{\prime\prime}$)} }

 \startdata 
1219$-$1154 & 12:19:34.88 & $-$11:54:22.9 & ... & ... & 60\tablenotemark{1} & 1.3 & ... & ...\\
1219$-$1201 & 12:19:44.49 & $-$12:01:35.5 & ... & ... & 60\tablenotemark{1} & 1.3 & ... & ...\\
1221$-$1206 & 12:21:46.20 & $-$12:06:12.7 & ... & ... & 45\tablenotemark{1} & 1.2 & ... & ...\\
1230+4621 & 12:30:16.26 & +46:21:17.1 & 67\tablenotemark{2} & 3.8 & 33\tablenotemark{2} & 2.2 & 72\tablenotemark{2} & 2.0\\
1326$-$1218 & 13:26:12.66 & $-$12:18:22.5 & 40\tablenotemark{3} & 0.9 & 50\tablenotemark{3} & 0.9 & 36\tablenotemark{3} & 0.8\\
1327$-$1217 & 13:27:57.33 & $-$12:17:16.7 & 20\tablenotemark{3} & 1.1 & 20\tablenotemark{3} & 1.0 & 42\tablenotemark{3} & 0.8\\
... & ... & ... & 40\tablenotemark{3} & 1.3 & 50\tablenotemark{3} & 0.8 & ... & ...\\
1329$-$1256 & 13:29:11.37 & $-$12:56:22.0 & 90\tablenotemark{1} & 1.6 & 95\tablenotemark{1} & 1.4 & 20\tablenotemark{4} & 2.0\\
... & ... & ... & 45\tablenotemark{3} & 1.1 & 30\tablenotemark{3} & 1.0 & ... & ...\\
1333$-$1237 & 13:33:01.92 & $-$12:37:17.0 & 20\tablenotemark{3} & 0.9 & 30\tablenotemark{3} & 1.2 & ... & ...\\
1404$-$1216 & 14:04:47.20 & $-$12:16:21.4 & 90\tablenotemark{1} & 1.5 & 60\tablenotemark{1} & 1.3 & ... & ...\\
1405$-$1147 & 14:05:11.38 & $-$11:47:08.6 & ... & ... & 60\tablenotemark{1} & 1.4 & ... & ...\\
1406$-$1232 & 14:06:36.54 & $-$12:32:39.7 & ... & ... & 45\tablenotemark{1} & 1.4 & ... & ...\\
1408$-$1209 & 14:08:17.86 & $-$12:09:27.0 & ... & ... & 68\tablenotemark{1} & 1.0 & ... & ...\\
1408$-$1216 & 14:08:45.95 & $-$12:16:08.8 & ... & ... & 68\tablenotemark{1} & 1.0 & ... & ...\\
1408$-$1218 & 14:08:50.62 & $-$12:18:14.8 & ... & ... & 68\tablenotemark{1} & 1.0 & ... & ...\\
1412$-$1150 & 14:12:32.56 & $-$11:50:16.2 & ... & ... & 8\tablenotemark{1} & 1.2 & ... & ...\\
... & ... & ... & ... & ... & 30\tablenotemark{3} & 0.8 & ... & ...\\
1412$-$1222 & 14:12:25.92 & $-$12:22:53.3 & ... & ... & 8\tablenotemark{1} & 1.3 & ... & ...\\
... & ... & ... & ... & ... & 45\tablenotemark{3} & 0.8 & ... & ...\\
1412$-$1222.1 & 14:12:31.26 & $-$12:22:15.8 & ... & ... & 45\tablenotemark{3} & 0.8 & ... & ...\\
1413$-$1244 & 14:13:08.59 & $-$12:44:11.6 & ... & ... & 40\tablenotemark{1} & 1.2 & ... & ...\\
1416$-$1143 & 14:16:45.13 & $-$11:43:40.0 & ... & ... & 40\tablenotemark{1} & 1.3 & ... & ...\\
1422+4622 & 14:22:24.18 & +46:22:39.7 & 80\tablenotemark{2} & 1.7 & 83\tablenotemark{2} & 2.4 & 35\tablenotemark{4} & 2.0\\
1519+4622 & 15:19:54.27 & +46:22:20.4 & ... & ... & 15\tablenotemark{2} & 2.3 & 45\tablenotemark{4} & 2.0
\enddata
\tablenotetext{(1)}{Las Campanas 1m}
\tablenotetext{(2)}{Palomar 1.5m}
\tablenotetext{(3)}{Las Campanas 2.5m}
\tablenotetext{(4)}{Lick 3m}
\end{deluxetable}
%\end{document}

\clearpage
%\documentclass[preprint]{aastex}
%\pagestyle{empty}
%\def\etal{et al.}
%\def\Kp{$K^\prime$\ }
%\begin{document}
\begin{deluxetable}{ccccccccc}
%\tabletypesize{\small}%\tablewidth{0pc}
\tablenum{2}
\tablecaption{BCG Data}
\tablewidth{340.22502pt}
\setlength{\tabcolsep}{0.06in}
\tablehead{
\colhead{Cluster} & \colhead{$z$} & \colhead{$\Sigma(1+z)^{5}$} & \colhead{m$_{V}$} & \colhead{m$_{I}$} & \colhead{m$_{K^{\prime}}$} & \colhead{$V-I$} & \colhead{$I-$\Kp} }

 \startdata
0915$+$4738 & 0.40 & ... & 20.26 & 18.44 & 15.37 & 1.83 & 3.07\\ 
0936$+$4620 & 0.54 & ... & 21.11 & 19.28 & 16.37 & 1.83 & 3.16\\
0944$+$4732 & 0.58 & ... & ... & ... & 16.49 & ... & ...\\
1002$-$1245 & 0.52 & 0.049 & 20.87 & 19.11 & ... & 1.76 & ...\\
1002$-$1247 & 0.64 & 0.090 & 21.05 & 18.72 & ... & 2.33 & ...\\
1005$-$1209 & 0.45 & ... & ... & 18.41 & ... & ... & ...\\
1007$-$1208 & 0.49 & 0.049 & ... & 18.02 & ... & ... & ...\\
1012$-$1243 & 0.49 & 0.039 & ... & 18.66 & ... & ... & ...\\
1012$-$1245 & 0.57 & ... & 21.82 & 19.16 & ... & 2.65 & ...\\
1014$-$1143 & 0.30 & ... & ... & 17.21 & ... & ... & ...\\
1015$-$1132 & 0.60 & 0.050 & ... & 18.54 & ... & ... & ...\\
1017$-$1128 & 0.50 & 0.117 & ... & 18.95 & ... & ... & ...\\
1018$-$1211 & 0.45 & 0.073 & 20.35 & 18.06 & ... & 2.29 & ...\\
1023$-$1303 & 0.51 & 0.125 & ... & 18.31 & ... & ... & ...\\
1024$-$1239 & 0.56 & 0.117 & 22.44 & 19.95 & ... & 2.49 & ...\\
1025$-$1236 & 0.36 & 0.071 & 18.84 & 16.97 & ... & 1.86 & ...\\
1027$-$1159 & 0.34 & 0.056 & 19.24 & 17.59 & ... & 1.65 & ...\\
1031$-$1244 & 0.68 & 0.138 & 21.20 & 18.64 & ... & 2.56 & ...\\
1032$-$1229 & 0.46 & 0.058 & 19.97 & 17.94 & ... & 2.03 & ...\\
1041$+$4626 & 0.62 & ... & ... & 18.64 & 15.73 & 2.91 & 2.91\\
1059$+$4737 & 0.36 & ... & 19.04 & 17.65 & 15.30 & 1.39 & 3.41\\
1100$+$4620 & 0.46 & ... & ... & ... & 15.69 & ... & ...\\
1136$-$1136 & 0.36 & 0.039 & 20.38 & 18.42 & ... & 1.96 & ...\\
1136$-$1145 & 0.65 & 0.101 & 21.54 & 18.84 & 15.72 & 2.70 & 3.12\\
1136$-$1252 & 0.36 & 0.056 & 19.63 & 17.63 & 14.91 & 2.00 & ...\\
1138$-$1142 & 0.45 & 0.058 & 20.40 & 18.20 & 16.19 & 2.20 & 2.44\\
1138$-$1225 & 0.33 & 0.094 & 18.11 & 16.73 & ... & 1.38 & ...\\
1138$-$1228 & 0.57 & ... & 21.24 & 19.10 & ... & 2.14 & ...\\
1139$-$1154 & 0.42 & ... & 20.95 & 19.10 & ... & 1.85 & ...\\
1139$-$1217 & 0.48 & ... & 20.29 & 17.96 & 15.31 & 2.33 & 2.65\\
1145$-$1155 & 0.48 & 0.047 & 20.46 & 18.32 & 15.73 & 2.13 & ...\\
1147$-$1252 & 0.58 & 0.074 & 21.71 & 19.13 & 16.48 & 2.58 & 2.73\\
1149$-$1159 & 0.32 & 0.073 & 18.49 & 17.05 & ... & 1.44 & ...\\
1149$-$1246 & 0.48 & 0.081 & 20.36 & 18.06 & 15.22 & 2.29 & 2.84\\
1208$-$1151 & 0.61 & ... & 21.10 & 18.59 & ... & 2.52 & ...\\
1210$-$1219 & 0.74 & 0.151 & 22.14 & 19.54 & ... & 2.60 & ...\\
1215$-$1252 & 0.39 & 0.063 & ... & 18.82 & ... & ... & ...\\
1216$-$1201 & 0.80 & 0.167 & ... & 18.94 & ... & ... & ...\\
1219$-$1201 & 0.56 & 0.069 & ... & 18.87 & ... & ... & ...\\
1221$-$1206 & 0.34 & 0.072 & ... & 17.49 & ... & ... & ...\\
1230$+$4621 & 0.51 & ... & 21.84 & 19.67 & 16.62 & 2.17 & 2.97\\
1326$-$1218 & 0.53 & ... & 20.99 & 18.80 & 17.00 & 2.19 & 2.61\\
1327$-$1217 & 0.44 & ... & 20.14 & 18.18 & 16.00 & 1.95 & 2.47\\
1329$-$1256 & 0.34 & 0.102 & 19.09 & 17.45 & 15.44 & 1.65 & 2.01\\
1333$-$1237 & 0.88 & ... & 22.54 & 19.50 & ... & 3.04 & ...\\
1404$-$1216 & 0.39 & 0.104 & 19.53 & 17.58 & ... & 1.96 & ...\\
1408$-$1209 & 0.34 & 0.051 & ... & 18.00 & ... & ... & ...\\
1408$-$1216 & 0.60 & 0.093 & ... & 19.09 & ... & ... & ...\\
1408$-$1218 & 0.60 & 0.186 & ... & 19.10 & ... & ... & ...\\
1412$-$1222 & 0.31 & ... & ... & 17.82 & ... & ... & ...\\
1412$-$1222.1 & 0.58 & ... & ... & 18.83 & ... & ... & ...\\
1416$-$1143 & 0.54 & 0.080 & ... & 18.21 & ... & ... & ...\\
1422$+$4622 & 0.69 & ... & 22.12 & 19.81 & 16.82 & 2.30 & 3.22\\
1519$+$4622 & 0.82 & ... & ... & 19.72 & 16.59 & ... & ...
\enddata
\end{deluxetable}
%\end{document}


\noindent

\begin{thebibliography}{}
\bibitem[a93]{a93}Aragon-Salamanca, A., Ellis, R.S., Couch, W.J., \& Carter,
D. 1993, \mnras, 262, 764, (A93)
\bibitem[a98]{a98}Aragon-Salamanca, A., Baugh, C.M., \& Kauffmann, G. 1998, \mnras, 297, 427 (ABK)
\bibitem[b96]{b96}Bertin, E, \& Arnouts, S. 1996, A\&AS, 117, 393
\bibitem[b92]{b92}Bower, R.G., Lucey, J.R., \& Ellis, R.S. 1992,
\mnras, 254, 601
\bibitem[b00]{b00}Burke, D.J., Collins, C.A., \& Mann, R.G. 2000,
\apj, 532L, 105 (BCM)
\bibitem[b93]{b93}Bruzual, G., \& Charlot, S. 1993, \apj, 405, 538
\bibitem[c98]{c98}Collins, C.A. \& Mann, R.G. 1998, \mnras, 297, 128
(CM)
\bibitem[d95]{d95}Dalcanton, J.J. 1995, Ph.D. Thesis, Princeton University
\bibitem[d97]{d97}Dalcanton, J.J., Spergel, D.N., Gunn, J.E., Schmidt,
M., \& Schneider, D.P. 1997, \aj, 114, 635
\bibitem[d96]{d96}van Dokkum, P.G. \& Franx, M. 1996, \mnras, 281, 985
\bibitem[d96]{d96}van Dokkum, P.G. \& Franx, M. 2001, \apj, 553, 90
\bibitem[d78]{d78}Dressler, A. 1978, \apj, 223, 765
\bibitem[g00]{g00}Gonzalez, A., Zaritsky, D., Dalcanton, J., \& Nelson, A. 2001a, in press
\bibitem[g00]{g00}Gonzalez, A., Zaritsky, D., Dalcanton, J., \&
Nelson, A. 2001b, submitted
\bibitem[g96]{g96}Graham, A., Lauer, T.R., Colless, M., \& Postman,
M. 1996 \apj, 465, 534
\bibitem[g87]{g87}Gunn, J.E. \etal 1987, Opt.Eng., 26, 779
\bibitem[h80]{h80}Hoessel, J.G. 1980, \apj, 241, 493
\bibitem[h80]{h80}Hoessel, J.G., Gunn, J.E., \& Thuan, T.X. 1980, \apj, 241, 486
\bibitem[h97]{h97}Hudson, M.J. \& Ebeling, H. 1997, \apj, 479, 621
\bibitem[h56]{h56}Humason, M.L., Mayall, N.U., \& Sandage, A.R. 1956,
\aj, 61, 97
\bibitem[k92]{k92}Kaastra, J.S., Asaoka, I., Koyama, K., \& Yamauchi,
S. 1992, \aa, 264, 654
\bibitem[k93]{k93}Kauffmann, G., White, S.D.M., \& Guiderdoni 1993,
\mnras, 264, 201
\bibitem[k95]{k95}Kauffmann, G. 1995a, \mnras, 274, 153
\bibitem[k95]{k95}Kauffmann, G. 1995b, \mnras, 274, 161
\bibitem[k00]{k00}Kelson, D.D., Illingworth, G.D., van Dokkum, P.G.,
\& Franx, M. 2000, \apj, 531, 159
\bibitem[k01]{k01}Kelson, D.D., Illingworth, G.D., Franx, M.,
\& van Dokkum, P.G., 2001, \apj, 552, 17 
\bibitem[k98]{k98}Kodama, T., Arimoto, N., Barger, A.J., \& Aragon-Salamanca, A. 1998, A\&A, 334, 99
\bibitem[l83]{l83}Landolt, A.U. 1983, \aj, 88, 439
\bibitem[l92]{l92}Landolt, A.U. 1992, \aj, 104, 340
\bibitem[l94]{l94}Lauer, T.R., \& Postman, M. 1994, \apj, 425, 418
\bibitem[l96]{l96}Lubin, L.M. 1996, \aj, 112, 23
\bibitem[m85]{m85}Mewe, R., Gronenschild, E.H.B.M., \& van den Oord, G.H.J. 1985, A\&AS, 62, 197
\bibitem[m85]{m86}Mewe, R., Lemen, J.R., \& van den Oord, G.H.J. 1986, A\&AS, 65, 511
\bibitem[m99]{m99}Mulchaey, J.S., \& Zabludoff, A.I. 1999, \apj, 514, 133
\bibitem[n00]{n00}Nelson, A.E., Gonzalez, A.H., Zaritsky, Z., \& Dalcanton, J.J. 2001a, in press
\bibitem[n00]{n00}Nelson, A.E., Simard, L., Zaritsky, Z., Dalcanton, J.J., \& Gonzalez, A.H. 2001b submitted
\bibitem[o76]{o76}Oemler, A. 1976, \apj, 209, 693
\bibitem[o95]{o95} Oke, J., \etal 1995, \procspie, 2198, 178 
\bibitem[p98]{p98}Persson, S.E., Murphy, D.C., Krzeminski, W., Roth, M., \& Riecke, M.J. 1998, \aj, 116, 2475
\bibitem[p95]{p95}Postman, M, \& Lauer, T.R. 1995, \apj, 440, 28 (PL95)
\bibitem[s76]{s76}Sandage, A., Kristian, J., \&  Westphal, J.A. 1976, \apj, 205, 688 
\bibitem[s96]{s96}Schetman, S.A., Landy, S.D., Oemler, A., Tucker, D.L., Lin, H., Kirshner, R.P., \& Schecter, P.L. 1996, \apj, 470, 172
\bibitem[s98]{s98}Schlegel, D.L., Finkbeiner, D.P., \& Davis, M. 1998, \apj, 500, 525
\bibitem[s94]{s94}Schneider, D.P., Schmidt, M., \& Gunn, J.E. 1994, \aj, 107, 124
\bibitem[s86]{s86}Schombert, J.M. 1986 \apjs, 60, 603
\bibitem[s97]{s97}Smail, I., Dressler, Al, Couch, W.J., Ellis, R.S., Oemler, A., Butcher, H., \& Sharples, R.M. 1997, \apjs, 100, 213
\bibitem[s98]{s98}Standard, S.A., Eisenhardt, P.R., \& Dickinson, M. 1998, \apj, 492, 461
\bibitem[s95]{s95}Stanford, S.A., Eisenhardt, P.R. \& Dickinson,
M. 1995, \apj, 450, 512
\bibitem[t77]{t77}Tremaine, S.D. \& Richstone, D.O. 1977, \apj, 212 311
\bibitem[v99]{v99}van Dokkum, P.G., Franx, M., Fabricant, D., Kelson, D.D.,
\& Illingworth, G.D. 1999, \apjl, 520, L95
\bibitem[z98]{z98}Zabludoff, A.I. \& Mulchaey, J.S. 1998, \apj, 498L, 5 
\bibitem[z97]{z97}Zaritsky, D., Nelson, A.E., Dalcanton, J.J., \& Gonzalez,
A.H. 1997, \apj, 480, L91 
\bibitem[z96]{z96}Zaritsky, D., Shectman, S.A., \& Bredthauer, G. 1996, \pasp, 108, 104
\end{thebibliography}
\end{document}